\documentclass[prx,twocolumn,amsmath]{revtex4}

\usepackage[utf8]{inputenc}
\DeclareUnicodeCharacter{2212}{$-$}
\usepackage{siunitx}
\usepackage{tabularx}
\usepackage{nccmath}
\usepackage{upgreek}
\usepackage{setspace}
\usepackage{amssymb}
\usepackage{mathtools}
\usepackage{tabu}
\usepackage{listings}
\usepackage{titlesec}
\usepackage{graphicx,natbib,mathrsfs,amsmath,amsbsy,bm,optidef}
\usepackage{optidef}
\usepackage[dvipsnames]{xcolor}
\definecolor{myblue}{named}{MidnightBlue}
\usepackage[colorlinks=true,citecolor=myblue,linkcolor=myblue,urlcolor=myblue]{hyperref}
\usepackage[caption=false]{subfig}
\usepackage{tabularx}

\begin{document}

\title{Responsive Operations for Key Services (ROKS): A Modular, Low SWaP Quantum Communications Payload}

\author{Craig D. Colquhoun}
\email{craig.colquhoun@craftprospect.com}
\affiliation{Craft Prospect Ltd, Suite 12 Fairfield, 1048 Govan Road, Glasgow, UK, G51 4XS}

\author{Hazel Jeffrey}
\affiliation{Craft Prospect Ltd, Suite 12 Fairfield, 1048 Govan Road, Glasgow, UK, G51 4XS}

\author{Steve Greenland}
\affiliation{Craft Prospect Ltd, Suite 12 Fairfield, 1048 Govan Road, Glasgow, UK, G51 4XS}

\author{Sonali Mohapatra}
\affiliation{Craft Prospect Ltd, Suite 12 Fairfield, 1048 Govan Road, Glasgow, UK, G51 4XS}

\author{Colin Aitken}
\affiliation{Craft Prospect Ltd, Suite 12 Fairfield, 1048 Govan Road, Glasgow, UK, G51 4XS}

\author{Mikulas Cebecauer}
\affiliation{Craft Prospect Ltd, Suite 12 Fairfield, 1048 Govan Road, Glasgow, UK, G51 4XS}

\author{Charlotte Crawshaw}
\affiliation{Craft Prospect Ltd, Suite 12 Fairfield, 1048 Govan Road, Glasgow, UK, G51 4XS}

\author{Kenny Jeffrey}
\affiliation{Craft Prospect Ltd, Suite 12 Fairfield, 1048 Govan Road, Glasgow, UK, G51 4XS}

\author{Toby Jeffreys}
\affiliation{Craft Prospect Ltd, Suite 12 Fairfield, 1048 Govan Road, Glasgow, UK, G51 4XS}

\author{Philippos Karagiannakis}
\affiliation{Craft Prospect Ltd, Suite 12 Fairfield, 1048 Govan Road, Glasgow, UK, G51 4XS}

\author{Ahren McTaggart}
\affiliation{Craft Prospect Ltd, Suite 12 Fairfield, 1048 Govan Road, Glasgow, UK, G51 4XS}

\author{Caitlin Stark}
\affiliation{Craft Prospect Ltd, Suite 12 Fairfield, 1048 Govan Road, Glasgow, UK, G51 4XS}

\author{Jack Wood}
\affiliation{Craft Prospect Ltd, Suite 12 Fairfield, 1048 Govan Road, Glasgow, UK, G51 4XS}

\author{Siddarth K. Joshi}
\email{sk.joshi@bristol.ac.uk}
\affiliation{Quantum Engineering Technology Labs \& Department of Electrical and Electronic Engineering, University of Bristol, UK}

\author{Jaya Sagar}
\affiliation{Quantum Engineering Technology Labs \& Department of Electrical and Electronic Engineering, University of Bristol, UK}

\author{Elliott Hastings}
\affiliation{Quantum Engineering Technology Labs \& Department of Electrical and Electronic Engineering, University of Bristol, UK}

\author{Peide Zhang}
\affiliation{Quantum Engineering Technology Labs \& Department of Electrical and Electronic Engineering, University of Bristol, UK}

\author{Milan Stefko}
\affiliation{Quantum Engineering Technology Labs \& Department of Electrical and Electronic Engineering, University of Bristol, UK}

\author{David Lowndes}
\affiliation{Quantum Engineering Technology Labs \& Department of Electrical and Electronic Engineering, University of Bristol, UK}

\author{John G. Rarity}
\affiliation{Quantum Engineering Technology Labs \& Department of Electrical and Electronic Engineering, University of Bristol, UK}

\author{Jasminder S. Sidhu}
\email{jasminder.sidhu@strath.ac.uk}
\affiliation{SUPA Department of Physics, University of Strathclyde, Glasgow, G4 0NG, UK}

\author{Thomas Brougham}
\affiliation{SUPA Department of Physics, University of Strathclyde, Glasgow, G4 0NG, UK}

\author{Duncan McArthur}
\affiliation{SUPA Department of Physics, University of Strathclyde, Glasgow, G4 0NG, UK}

\author{Roberto G. Pousa}
\affiliation{SUPA Department of Physics, University of Strathclyde, Glasgow, G4 0NG, UK}

\author{Daniel K. L. Oi}
\affiliation{SUPA Department of Physics, University of Strathclyde, Glasgow, G4 0NG, UK}

\author{Matthew Warden}
\email{matthew.warden@fraunhofer.co.uk}
\affiliation{Fraunhofer Centre for Applied Photonics, Technology and Innovation Centre, 99 George Street, Glasgow, G1 1RD, UK}

\author{Eilidh Johnston}
\affiliation{Fraunhofer Centre for Applied Photonics, Technology and Innovation Centre, 99 George Street, Glasgow, G1 1RD, UK}

\author{John Leck}
\affiliation{Fraunhofer Centre for Applied Photonics, Technology and Innovation Centre, 99 George Street, Glasgow, G1 1RD, UK}

\begin{abstract}
Quantum key distribution (QKD) is a theoretically proven future-proof secure encryption method that inherits its security from fundamental physical principles. With a proof-of-concept QKD payload having flown on the Micius satellite since 2016, efforts have intensified globally. Craft Prospect, working with a number of UK organisations, has been focused on miniaturising the technologies that enable QKD so that they may be used in smaller platforms including nanosatellites. The significant reduction of size, and therefore the cost of launching quantum communication technologies either on a dedicated platform or hosted as part of a larger optical communications will improve potential access to quantum encryption on a relatively quick timescale.

The Responsive Operations for Key Services (ROKS) mission seeks to be among the first to send a QKD payload on a CubeSat into low Earth orbit, demonstrating the capabilities of newly developed modular quantum technologies. The ROKS payload comprises a quantum source module that supplies photons randomly in any of four linear polarisation states fed from a quantum random number generator; an acquisition, pointing, and tracking system to fine-tune alignment of the quantum source beam with an optical ground station; an imager that will detect cloud cover autonomously; and an onboard computer that controls and monitors the other modules, which manages the payload and assures the overall performance and security of the system. Each of these modules have been developed with low Size, Weight and Power (SWaP) for CubeSats, but with interoperability in mind for other satellite form factors.

We present each of the listed components, together with the initial test results from our test bench and the performance of our protoflight models prior to initial integration with the 6U CubeSat platform systems. The completed ROKS payload will be ready for flight at the end of 2022, with various modular components already being baselined for flight and integrated into third party communication missions.
\end{abstract}

\maketitle

\section{INTRODUCTION}

\noindent
Responsive Operations for Key Services (ROKS) is a UK Space Agency funded mission that aims to launch a 6U CubeSat with quantum key distribution (QKD) and cloud-detection capabilities~\cite{FLIWhitePaper} into Low-Earth Orbit (LEO). ROKS has supported the progression of QKD instrumentation from apparatus spanning optical test benches in university laboratories, to miniaturised, modular, space-ready subsystems. 

Craft Prospect Ltd (CPL) and partners have designed, manufactured, assembled, tested, and now integrated several modules for this mission~\cite{CassSmallsatPaper}, though with reconfigurability and interoperability in mind for future flight opportunities. This approach enables ROKS subsystems to be supplied either individually or bundled for a range of satellite form factors. The individual capabilities of one of CPL's modules will be demonstrated in the upcoming Canadian space agency mission QEYSSat~\cite{QEYSSATPaper} - a variant of the JADE quantum source produced for ROKS will be driven by a QRNG board developed at University of Waterloo as a secondary payload on the satellite, demonstrating its ability to interface with different driving electronics, and with different QRNGs~\cite{QRNGPaper} from a range of suppliers. 

The remainder of the paper is laid out as follows: the rest of this section gives the motivation for using orbital CubeSats for QKD, Section~\ref{sec:MissionPayload} lists and describes each of the modules and their purposes in the mission, Section~\ref{sec:ModuleTesting} presents some key module tests and results, Section~\ref{sec:OGS} describes some of the ongoing Optical Ground Station work at University of Bristol, some of the lessons learned as part of the test and integration process are listed in Section~\ref{sec:Lessons}, and Section~\ref{sec:Conclusions} concludes the paper. 

\subsection{Why QKD?}
\noindent
Encryption, the use of secret keys to encrypt or decrypt information, is now ubiquitous on the internet, to the point that most people use encryption on a daily basis. Two of the most common types of encryption used to secure data online are RSA~\cite{RSA} which relies on the multiplication of two large prime numbers for its key generation, and AES which involves performing several matrix operations to do the same. There are algorithms for quantum computers that threaten the security of these methods - Shor's algorithm~\cite{ShorPaper} significantly reduces the complexity of prime number factorisation when compared with classical computers, and Grover's algorithm~\cite{GroverPaper} effectively halves the bit length of keys for brute forcing attacks. The former poses a significant threat~\cite{NISTPostQuantumCrypto} to RSA encryption of all currently used bit lengths, and the latter greatly reduces the amount of time it would take to correctly guess an AES key assuming no vulnerabilities for the method would be discovered. 

With technological advances improving the scalability, fault tolerance, and commercial availability of quantum computers; it is only a matter of time until cyber criminals and state actors gain access to them, and until these encryption methods are rendered insecure. Such nefarious actors need not wait until capable quantum computers are available to decrypt confidential information, they may already be storing encrypted data to decrypt later, so quantum-secure encryption methods are critical at present~\cite{NISTPostQuantumCrypto}. Quantum key distribution (QKD) is an encryption method that relies on the quantum mechanical properties of light for its key generation rather than mathematical complexity, making it theoretically secure against attacks from quantum computers~\cite{QKDPrivacyNoisy, Sidhu2021advances}.

While other QKD protocols use the polarisations of entangled photons or time binning to distinguish between photons, or qubits, that represent a 0 or 1, in ROKS a symmetric BB84~\cite{cointossing} protocol is used. In the quantum communications channel, weak coherent pulses ($<$1 photon per pulse average) are generated in one of four linear polarisations at random: horizontal (H) $\mathrlap{\leftarrow}\rightarrow$, vertical (V) $\updownarrow$, diagonal (D) $\mathrlap{\swarrow}\hspace{-3.8pt}\nearrow$, or anti-diagonal (A) $\mathrlap{\searrow}\hspace{-3.8pt}\nwarrow$. These polarisations are paired off into two polarisation bases with their orthogonal counterparts - the H-V basis $\mathrlap{\leftarrow}\mathrlap{\hspace{4.5pt}\updownarrow}\rightarrow$, and the D-A basis $\mathrlap{\mathrlap{\swarrow}\hspace{-2.9pt}\nearrow}\mathrlap{\searrow}\hspace{-3.8pt}\nwarrow$ - within which one polarisation corresponds to a 1 and the other to a 0. All properties of the photons other than their polarisations should be identical so the photons are otherwise indistinguishable from one another.

After the bits have been sent a reconciliation process occurs over classical communication channels, during which only the basis in which each photon was sent is transmitted so the receiver can determine if the photon was detected in the correct polarisation basis. If the basis is correct, the qubit has almost certainly been correctly measured so it is kept; if the basis is incorrect, the qubit will only be correctly measured 50\% of the time, so this bit is ignored and does not contribute to the final secret key. Because only single photons are transmitted, any eavesdroppers can be detected by decreases in the numbers of detected photons, which can also be stated as an increase in the quantum bit error rate (QBER). Eavesdroppers also cannot reliably reproduce photons they pick off, because there is a 50\% probability they measure each photon in the incorrect polarisation basis. 

As well as `signal states' that are kept and contribute to the final secret key, `decoy states'~\cite{DecoyQKD} are also used for added complexity in case of eavesdroppers. Decoy states contain all the same information as signal states, just at a different intensity. During the reconciliation process over the classical communications channel, the transmitter declares which pulses were signal and which were decoy. 

Some BB84 QKD solutions are commercially available, either coupled into free-space or fibre-coupled. The distances over which these solutions can be used are limited by atmospheric losses~\cite{FreespaceQKD} and fibre attenuation~\cite{FibreQKD}, respectively. Using satellites from various Earth orbits should overcome these limits~\cite{CubesatQKD, Sidhu2020_npjQI, sidhu2021key} as this can offer global reach, and the atmosphere exists at relatively low altitudes for some of which its density is reduced, limiting atmospheric losses~\cite{QUARCPaper, Rarity_atmosphere}. The ROKS mission aims to demonstrate the feasibility of supplying quantum keys from CubeSats in LEO~\cite{Islam2022finite}, with the potential for constellations in future to reduce lapses in coverage at compatible ground stations~\cite{Gundogan2021npjQI, gundogan2021topical, Wallnofer2022_NC}. 

\subsection{MISSION GOALS}
\noindent
ROKS is a demonstrator mission both for the devices developed for the platform, and the services that these devices can provide. The main mission goals are to demonstrate that QKD technology can be successfully implemented on a CubeSat~\cite{Sidhu2020_npjQI, Satquma2021documentation}, and to demonstrate that onboard intelligence can be used to deliver mission autonomy and improve utility of a potential CubeSat QKD service. 

\section{MISSION PAYLOAD}
\label{sec:MissionPayload}
\begin{figure}
	\centering
	\includegraphics[width = \linewidth]{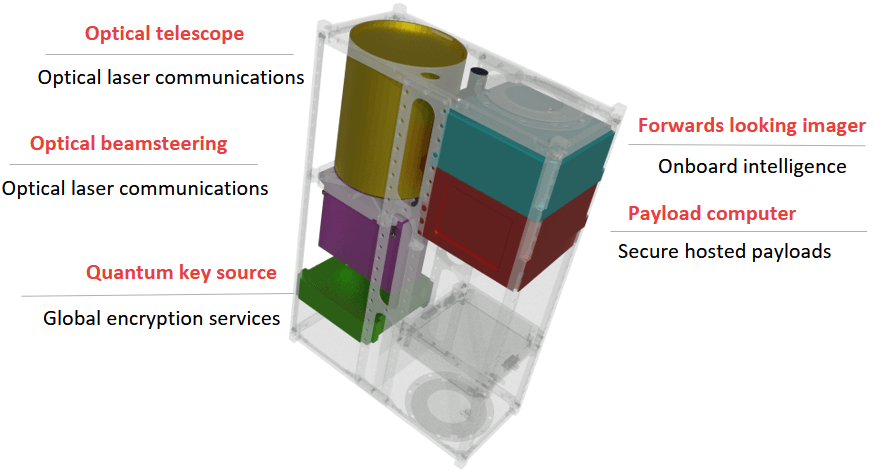}
	\caption{A Computer Render of the 6U Structure with Key Components Highlighted}
	\label{fig:ROKSRender}
\end{figure}

\noindent
The ROKS payload consists of 5 modular hardware subsystems, described below. The quantum source (JADE) and acquisition, pointing, and tracking (APATITE) components were designed and manufactured in conjunction with mission partners University of Bristol (UoB) and University of Strathclyde (UoS). The optical telescope (GARNET) was developed by Fraunhofer Centre for Applied Photonics (FhCAP), and licensed to CPL for commercialisation. Each of these modules is shown in its position in the CubeSat structure in Fig.~\ref{fig:ROKSRender}.

\begin{figure}
	\centering
	\includegraphics[width = \linewidth]{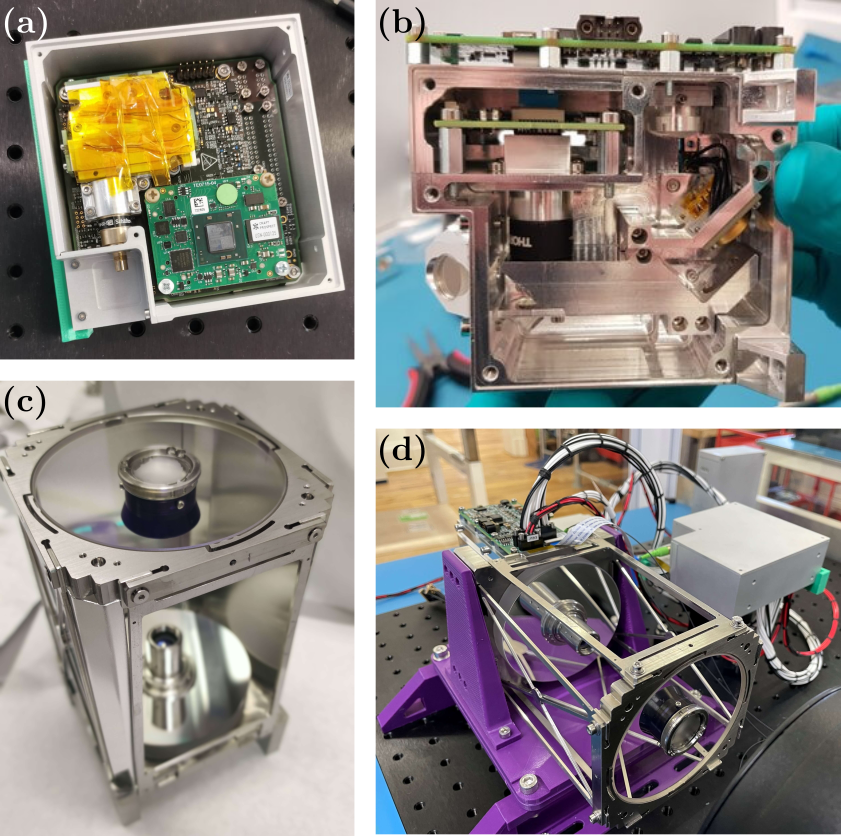}
	\caption{Images of the Modules Contained in the Optical Payload (OPAL) Segment of ROKS (a) JADE Quantum Source Module; (b) APATITE Acquisition, Pointing, and Tracking Module; (c) GARNET Optical Telescope; (d) All the OPAL Modules Connected in a Test Bench Configuration}
	\label{fig:OPALCombined}
\end{figure}

\noindent \underline{JADE}: The JADE quantum source combines optical components that produce quantum signal pulses (GNEISS), with the electronics used to drive them. The JADE PCB includes a Zynq FPGA, a series of quantum random number generators (QRNGs), several laser diode driver chips intended for pulsed operation, sensors to provide telemetry for the module, and interfaces to other modules. Operational flexibility is provided by the ability to tune FPGA parameters during runtime - the user may tune the pulse rate, individual pulse widths and laser currents, or whether to pulse using numbers form the QRNGs or to pulse arbitrary patterns for test and calibration purposes. The specified pulse rate for the ROKS mission is 100\,MHz, with 1\,ns FWHM pulse width, generating signal states of 0.8 photons per pulse at the telescope exit and two decoy states of 0.4 and 0 photons per pulse. An image of the JADE module internals is shown in Fig.~\ref{fig:OPALCombined}(a).

\noindent \underline{GNEISS}: The optical component of the JADE module generates weak coherent pulses of 785\,nm light in four linear polarisations. Thermal control of each laser diode is used to tweak the wavelengths for indistinguishability between photon sources. There is also a photodiode for diagnostic purposes, and an 830 nm alignment laser source that is used in the APT subsystem. Crucially, all the beams produced in the module are coupled into a single optical fibre so that they all share the same spatial mode upon leaving the JADE module. \vspace{10pt}

\noindent \underline{APATITE}: The APATITE acquisition, pointing, and tracking system uses a series of dichroic mirrors to separate the alignment beam from the quantum source beams generated by JADE, and combine the quantum source light with a downlink beacon laser beam. The alignment beam is directed onto a camera sensor as reference for where the quantum source is pointing - the camera sensor will also be used to detect uplink beacon light from compatible optical ground stations (OGSs). The downlink beacon beam serves three purposes: it provides an alignment signal for OGSs; it serves as a polarisation reference so the OGS may calibrate its optics to optimise detection of the quantum source polarisation states; and it supplies timing and synchronisation information during the quantum key transmission phase. The alignment of the downlink beacon, the quantum source, and the uplink beacon beams is managed using a MEMS mirror. The optomechanical parts, camera board, and APATITE driver board can be seen inside the enclosure in Fig.~\ref{fig:OPALCombined}(b). Any light that leaves APATITE has to pass through the satellite's telescope.

\noindent \underline{GARNET}: The optical telescope was developed by the Fraunhofer Centre for Applied Photonics (FhCAP) and licensed to CPL. This Schmidt-Cassegrain reflecting telescope was designed specifically for ROKS but it can be used for any optical communications CubeSat in principle. GARNET (shown in Fig.~\ref{fig:OPALCombined}(c)) occupies a volume of 1.5\,U, has a 90\,mm (80\,mm clear) aperture and provides 30x magnification to outgoing beams with low distortion with a $\pm$0.25$^\circ$ field of view. Its exit pupil is located externally to the telescope to allow convenient interfacing with a steering mirror - on ROKS it interfaces directly with the front of APATITE. The telescope housing is designed to minimise contact with the CubeSat platform to reduce environmental stresses experienced by the optics.

\noindent \underline{OPAL}: The optical payload (OPAL) encompasses JADE, APATITE, and GARNET as the full optical subsystem, as shown in Fig.~\ref{fig:OPALCombined}(d). This is defined as a bundle that may be offered for future commercial flight opportunities rather than its individual parts.

\begin{figure}
	\centering
	\includegraphics[width = 0.7\linewidth]{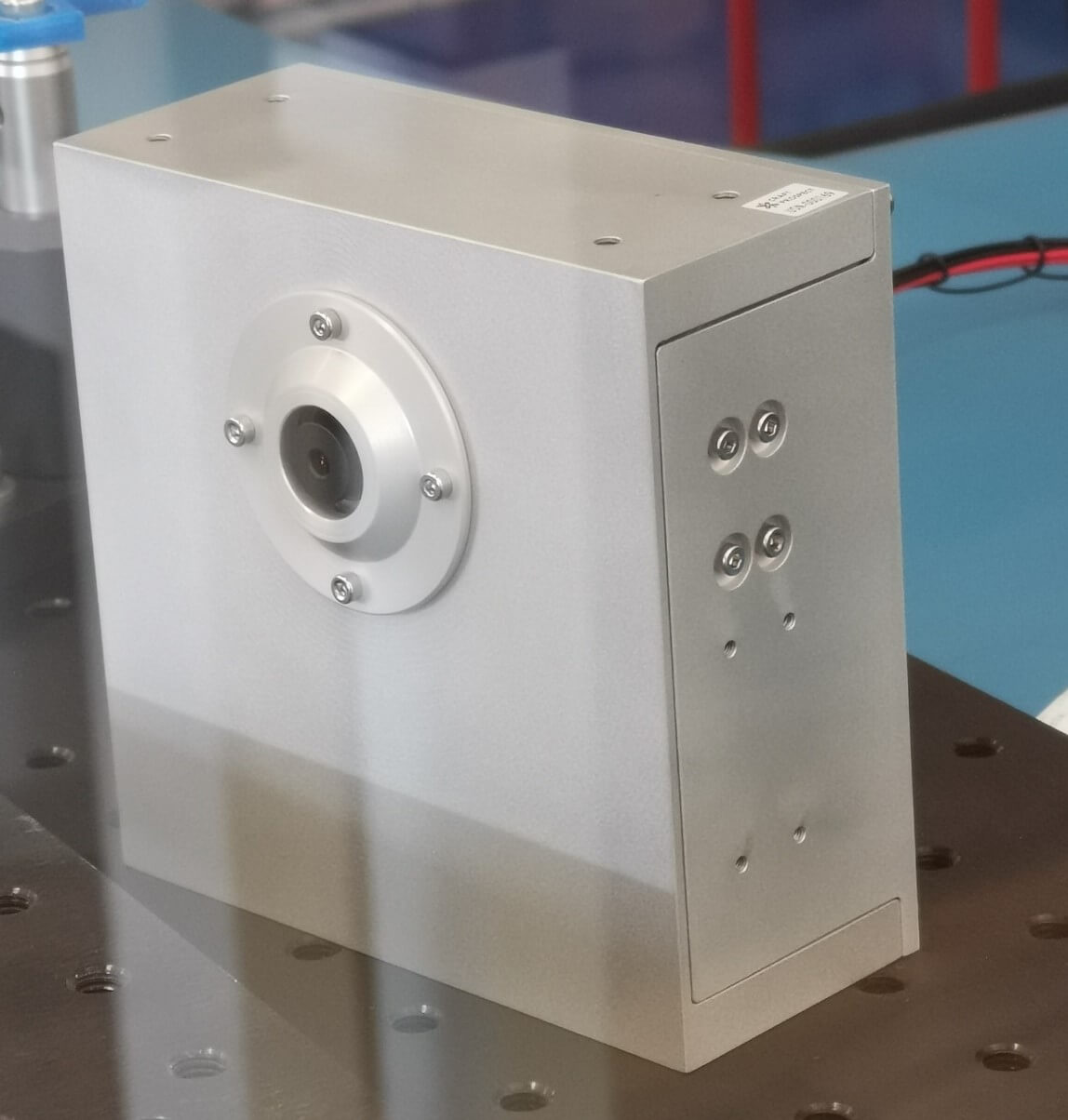}
	\caption{FLI-NT Module}
	\label{fig:FLINTExt}
\end{figure}

\noindent \underline{FLI-NT}: The Forwards Looking Imager (FLI), CPL's first commercially available product, is a camera sensor instrument powered by a Zynq FPGA embedded with a deep learning algorithm for user-defined EO feature classification. Expected use cases include wildfire monitoring, ship tracking, and, as in the case of ROKS, cloud detection. The FLI-NT name is given to the ROKS model of FLI because it has been designed and trained for the more challenging purpose of detecting clouds at night time (NT), on the dark side of Earth. The module `looks' ahead of the CubeSat's location, identifies whether clouds will block line-of-sight with the target OGS, and allows the onboard computer to either prepare for QKD execution, or to direct the CubeSat's resources elsewhere, generating more random numbers for the next pass for instance. FLI-NT (shown in Fig.~\ref{fig:FLINTExt}) is configurable to offer up to 120\,s look-ahead for the ROKS mission.

\begin{figure}
	\centering
	\includegraphics[width = 0.7\linewidth]{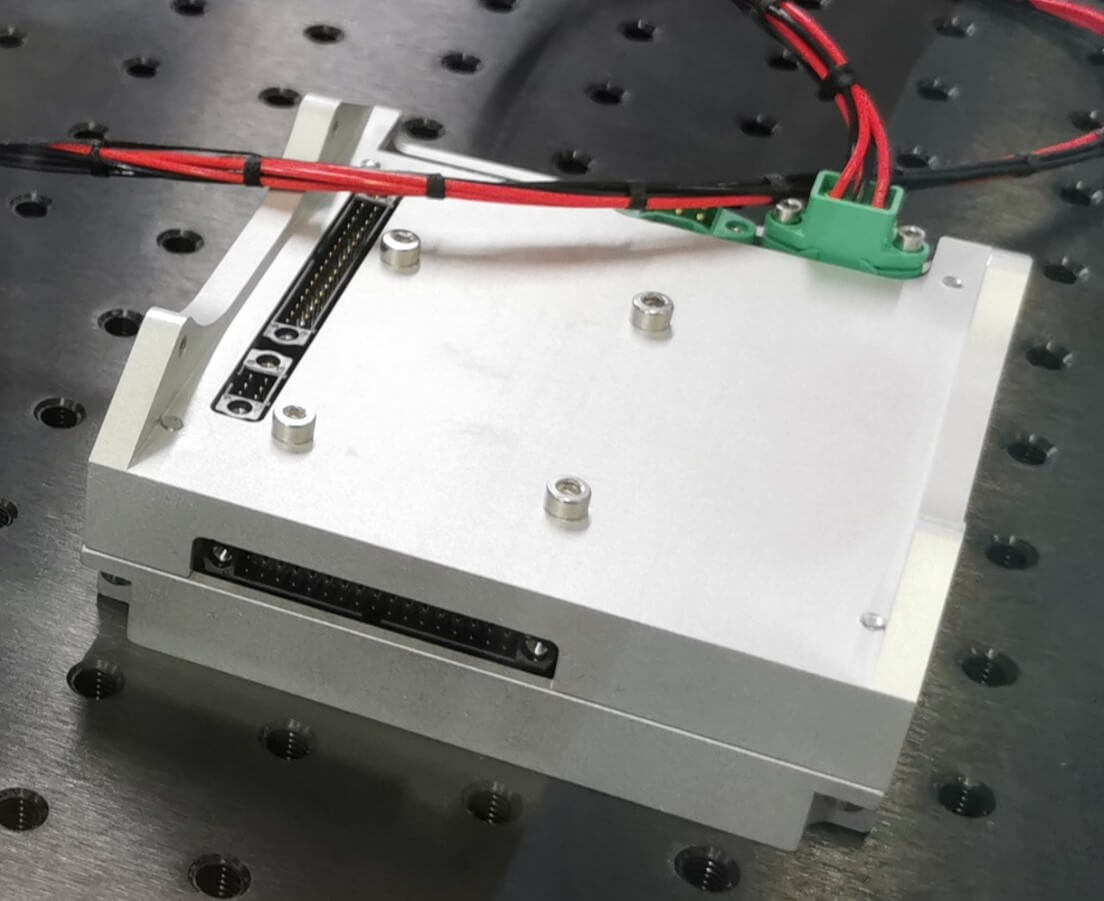}
	\caption{OBSIDIAN Onboard Computer}
	\label{fig:OBSIDIANExt}
\end{figure}

\noindent \underline{OBSIDIAN}: The OBSIDIAN onboard computer interfaces with all the previously described subsystems using a Zynq-based FPGA running Bright Ascension's GenOne flight software with CPL components for responsive scheduling and QKD. This module also has high-speed throughput to the rf transceiver for the reconciliation stage of the QKD process and telemetry transmissions. OBSIDIAN is shown in Fig.~\ref{fig:OBSIDIANExt}.

\section{MODULE TESTING}
\label{sec:ModuleTesting}
\noindent
Prior to the payload integration phase of the AIT process, modules were calibrated and tested on an individual basis. 

\subsection{JADE Quantum Source}

\begin{figure}
	\centering
	\includegraphics[width = \linewidth]{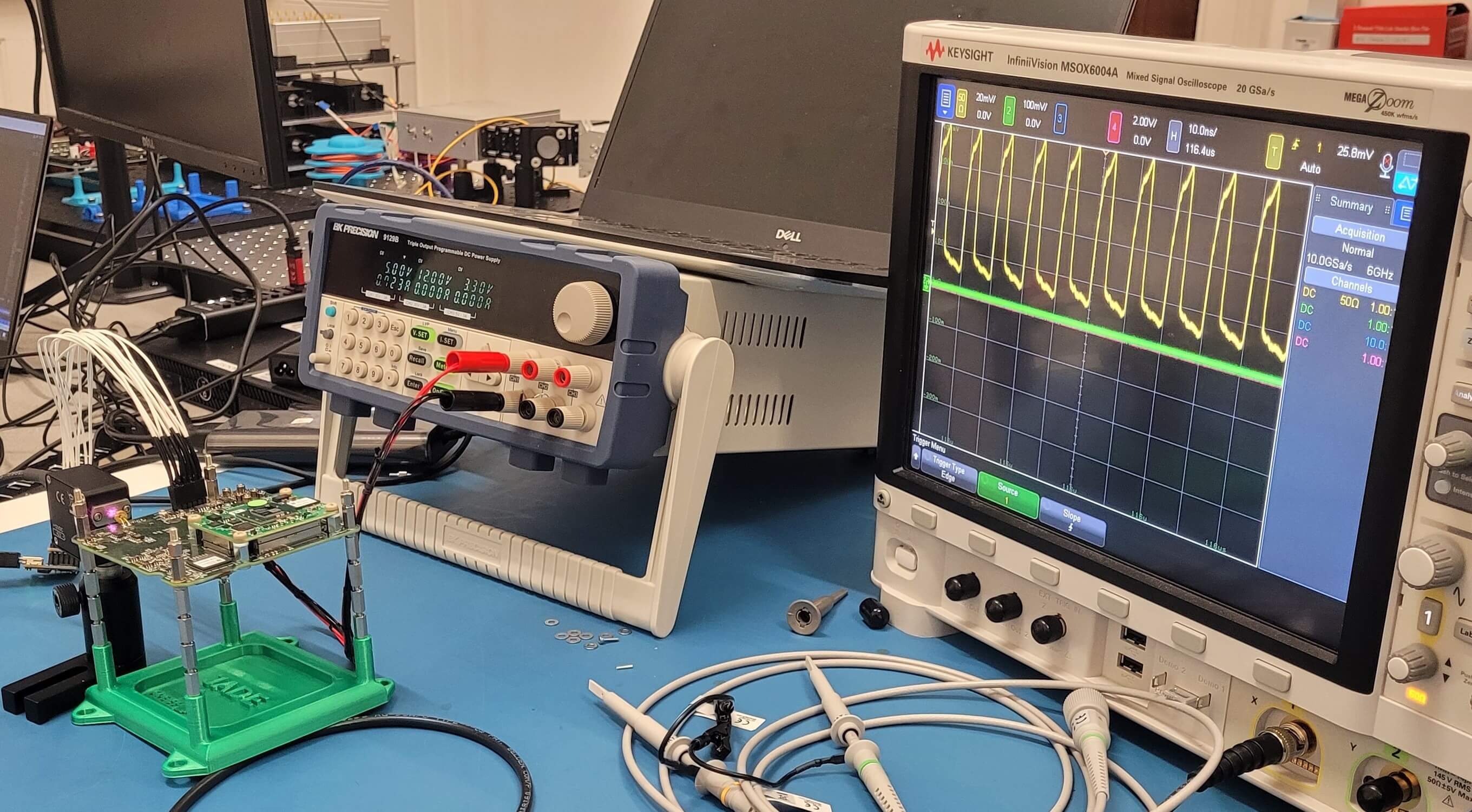}
	\caption{JADE Test Apparatus }
	\label{fig:JADEElectricalTest}
\end{figure}

\noindent \underline{Pulse test and calibration}: During JADE electronics board bring-up, various test points on the PCB and secured components are probed to confirm that the expected signals and voltages are being produced. One test that can efficiently identify if there are problems on the board is to examine optical pulses produced by bare laser diodes connected to the drivers, giving an end-to-end measure of the driving electronics and the firmware flashed onto the FPGA. A successful run of this test is shown in Fig.~\ref{fig:JADEElectricalTest}, where nominal 100\,MHz optical pulses that are representative of what will be generated by the JADE quantum source laser diodes can be observed on the oscilloscope screen.

\begin{figure}
	\centering
	\includegraphics[width = \linewidth]{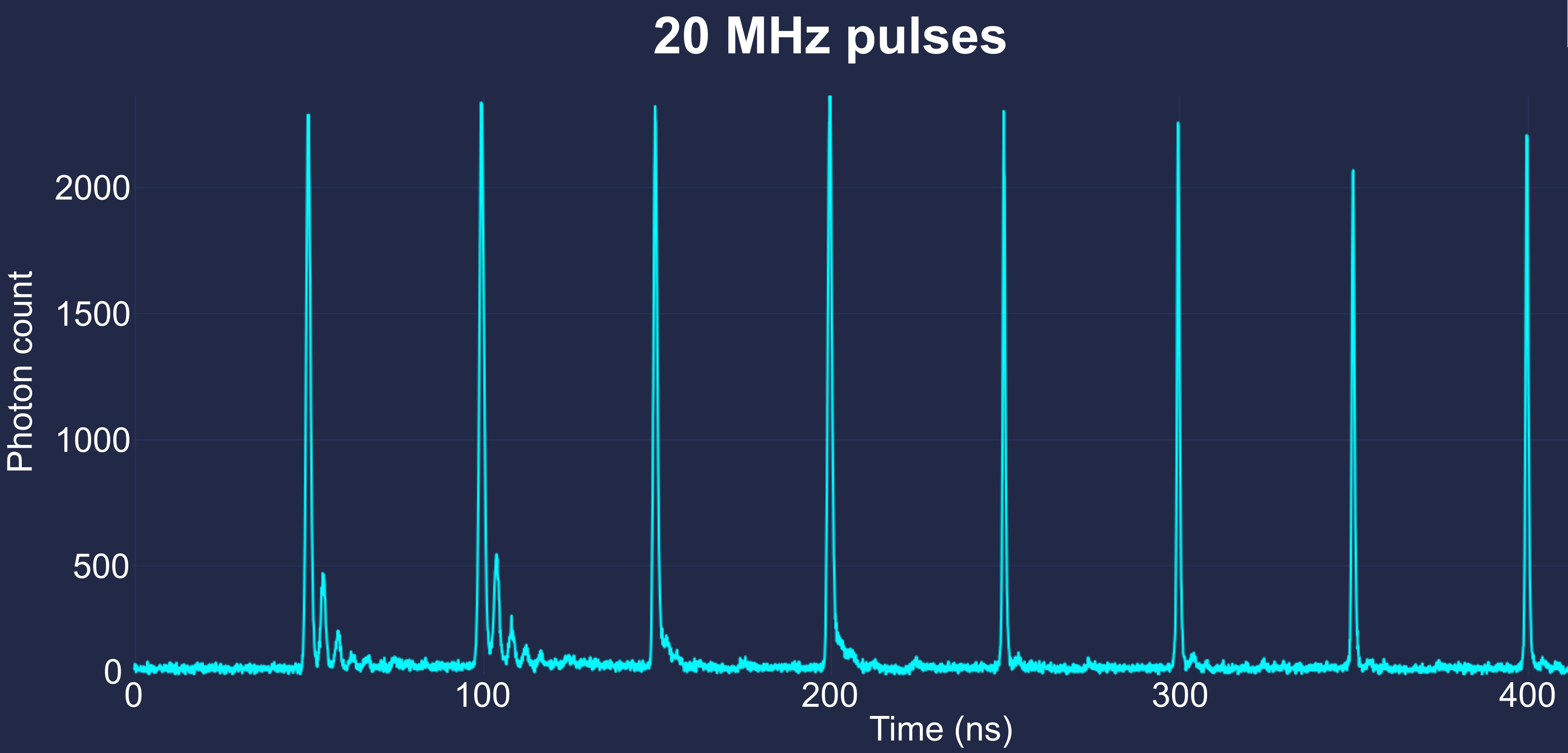}
	\caption{An Accumulative Histogram Showing JADE Pulses at 20\,MHz}
	\label{fig:JADEAccumulativeHist}
\end{figure}

The full system is tested once the JADE PCB and GNEISS are integrated. Although an effort is made to optimise the coupling efficiency of all the laser beams in GNEISS into the fibre, these efficiencies may fluctuate while the optical glue cures. The difference in coupling efficiencies is amended during the JADE calibration process by connecting the GNEISS fibre directly to a single photon counting module (SPCM) and timetagger system; pulsing all the laser diodes in a test pattern and tweaking the current of each laser until the photon numbers are equal. An accumulative histogram showing pulses in the test pattern at 20\,MHz is shown in Fig.~\ref{fig:JADEAccumulativeHist}. A modulation frequency of 20\,MHz is used to so that the signals are unaffected by SPCM dead time during calibration of the pulse heights or photon numbers. In this case an attempt was made to equalise the pulse heights. 

\noindent \underline{Polarisation testing}: Since polarisations are critical to BB84 QKD, polarisation testing occurs at several stages throughout the integration process to ensure polarisation-dependent effects are minimised. Polarisation rotations are normal and expected, but if anything in ROKS causes shifts or losses in some polarisations more than others this will increase the error rate of our QKD system. Ideally all polarisations should be separated by 45$^\circ$. Once GNEISS and JADE are integrated the quantum source laser diodes can only be pulsed, so the signal can not easily be detected by a polarimeter. 

\begin{figure}
	\centering
	\includegraphics[width = \linewidth]{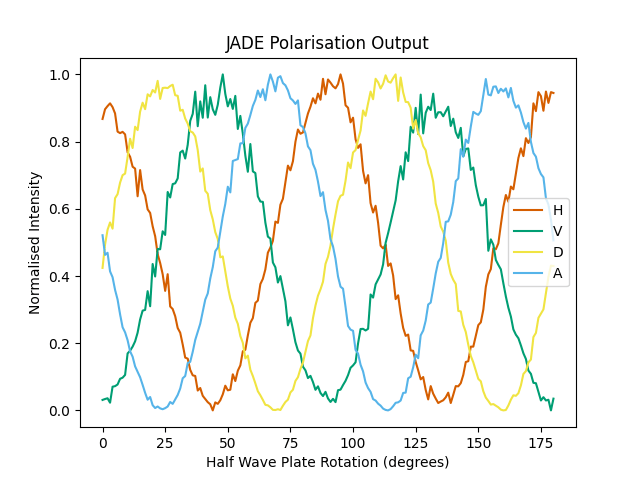}
	\caption{A Relative Polarisation Measurement Directly from JADE}
	\label{fig:JADERelativePolarisationMeasurement}
\end{figure}

CPL uses two different apparatus for polarisation testing - one for simple readout of the relative polarisations generated by JADE, and another to split the polarisation states into four different beam paths, closely resembling the instrument that will be used in ROKS-compatible OGSs. In both cases, two quarter wave plates (QWPs) are used to linearise the polarisation (as ellipticity can be introduced in the fibre and optics) and a half wave plate (HWP) is used to rotate that linear polarisation. For the relative polarisation test the wave plates are placed in front of a polarising beam splitter (PBS), the HWP is motorised, rotating continuously. An SPCM is placed on the reflected arm of the PBS, detecting changes of intensity as the HWP is rotated. Given a known test pattern, the H, V, D, and A photons can be identified in the signals generated in the detector. The photons counted in each pulse are plotted against HWP angle, as shown in Fig.~\ref{fig:JADERelativePolarisationMeasurement}. Here it can be seen that the polarisations are all separated by 22.5$^\circ$ of HWP rotation, corresponding to 45$^\circ$ polarisation separation. 

\begin{figure}
	\centering
	\includegraphics[width = \linewidth]{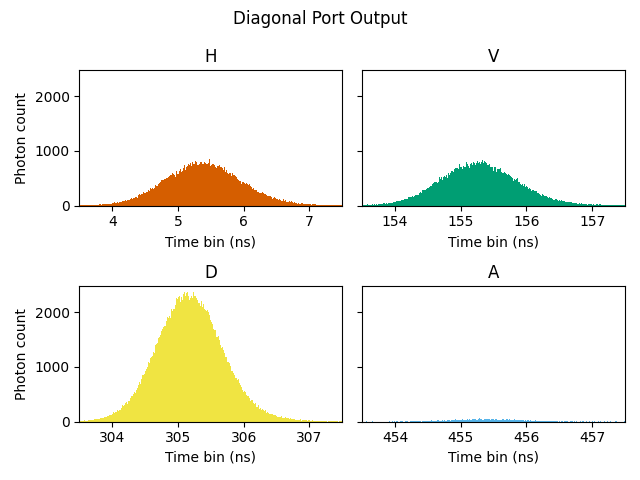}
	\caption{An Accumulative Histogram Result Obtained Using Polarisation Readout Apparatus}
	\label{fig:JADEPolarisation21}
\end{figure}

\begin{figure}
	\centering
	\includegraphics[width = \linewidth]{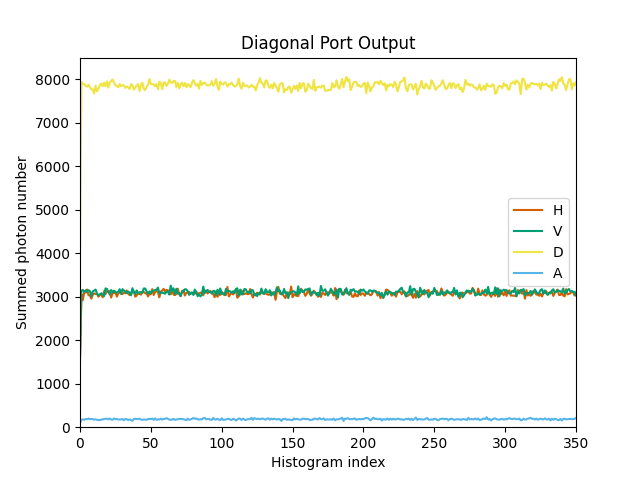}
	\caption{The Sums of the Histograms Shown in Fig.~\ref{fig:JADEPolarisation21}. The Lines for H and V Are Almost Completely Overlapped}
	\label{fig:JADEOscopeResult}
\end{figure}

The polarisation readout instrument is more elaborate, using a combination of PBSs and non-polarising BSs to direct the light to one of four SPCMs. In this case, the wave plates are set to angles that maximise the signal for a polarisation state at each detector, then locked in position. Figure~\ref{fig:JADEPolarisation21} presents four regions of interest (ROIs) on an accumulative histogram showing photon detection statistics from a test pattern. As these results were obtained from the port maximised for diagonal photon detection, it is expected that the D ROI should have the most photon counts, the A ROI should have almost no counts, and the H and V should have equal counts each approximately 50\% of the D count, coming from the opposite polarisation basis. Figure~\ref{fig:JADEOscopeResult} shows the sum of photons in each pulse against time, illuminating that these proportions are correct and that the system is working as expected.


\noindent \underline{Thermal testing}: Several thermal cycles are performed on each ROKS module to ensure that all functions work nominally throughout the operating temperature range (-20$^\circ$C - 50$^\circ$C); to check that the modules behave normally when hot and cold starts are performed at the extremes of the operating temperature range; and to test that the modules switch on and behave normally after being cycled through the survival temperature range (-30$^\circ$C - 80$^\circ$C). All ROKS modules passed this battery of thermal tests, key data was collected at the operating temperature extremes, and telemetry was collected throughout the tests where possible.

\begin{figure}
	\centering
	\includegraphics[width = \linewidth]{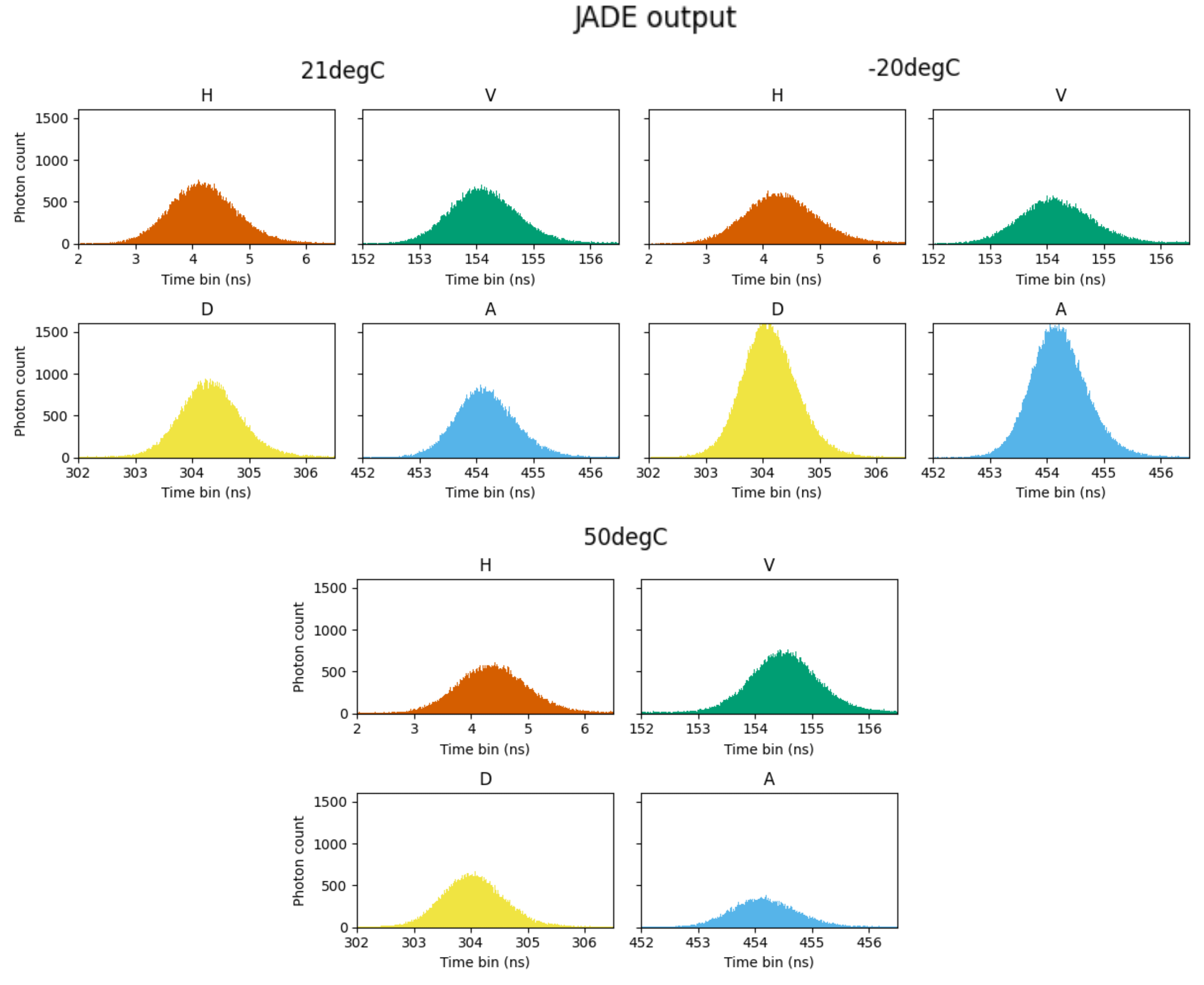}
	\caption{Accumulative Histogram Results for the Light Directly Output by JADE for 21$^\circ$C, -20$^\circ$C, and 50$^\circ$C}
	\label{fig:JADEThermal}
\end{figure}

\begin{figure}
	\centering
	\includegraphics[width = 0.8\linewidth]{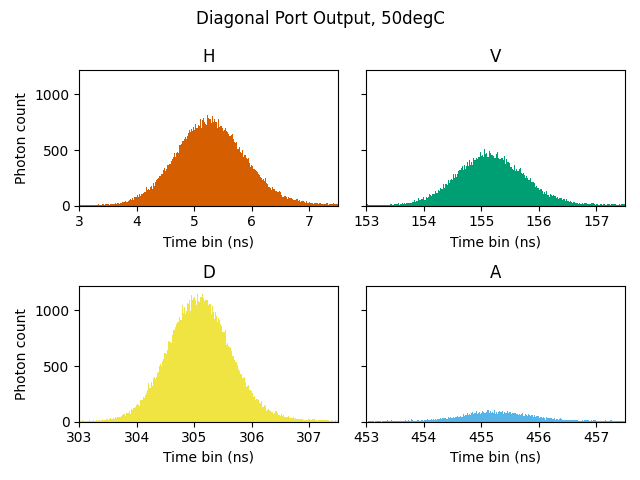}
	\caption{A Polarisation Readout Result Taken at 50$^\circ$C from the Same Port as in Fig.~\ref{fig:JADEPolarisation21}}
	\label{fig:JADEPolarisation50}
\end{figure}

The same JADE functional tests as above were performed at the operating temperature extremes. The photon output directly from JADE is shown in Fig.~\ref{fig:JADEThermal} for various temperatures - there is a clear change in fibre coupling efficiency, and perhaps the optical power produced by the laser diodes, at different temperatures. The most surprising observation from these results is that from room temperature to -20$^\circ$C, the number of photons detected from diodes D and A almost doubles. None of the observed drifts pose a threat to QKD security, the change in photon number vs temperature must be well characterised for the flight model so the laser diode modulation current can be adjusted based on JADE telemetry to tune the photons per pulse.
In addition to the fibre coupling fluctuations, the change in temperature also induces a change of birefringence in the fibre, leading to polarisation drifts. Polarisation drifts experienced by light traversing the fibre can be seen by comparing Fig.~\ref{fig:JADEPolarisation21} and Fig.~\ref{fig:JADEPolarisation50}. At 50$^\circ$C, there is now a difference in photon numbers between H and V, which previously had equal outputs through the diagonal port on the polarisation readout apparatus. In addition, an increase in the A signal can be observed which would lead to increased errors if not accounted for using the waveplates in front of the instrument. Again, this drift does not pose a problem for ROKS, it simply needs to be well characterised prior to launch, and can be corrected for in the ground station. Additionally, this test was performed using a 1\,m long optical fibre inside the thermal chamber and a 30\,cm long fibre will be used in flight, so this polarisation rotation will be less pronounced during the mission.

\subsection{APATITE APT Subsystem}

\noindent \underline{Camera sensor / alignment laser testing}: Camera sensor tests are often the first performed for an APATITE module because of how diagnostically useful the camera is for further testing. 

\begin{figure}
	\centering
	\includegraphics[width = 5cm]{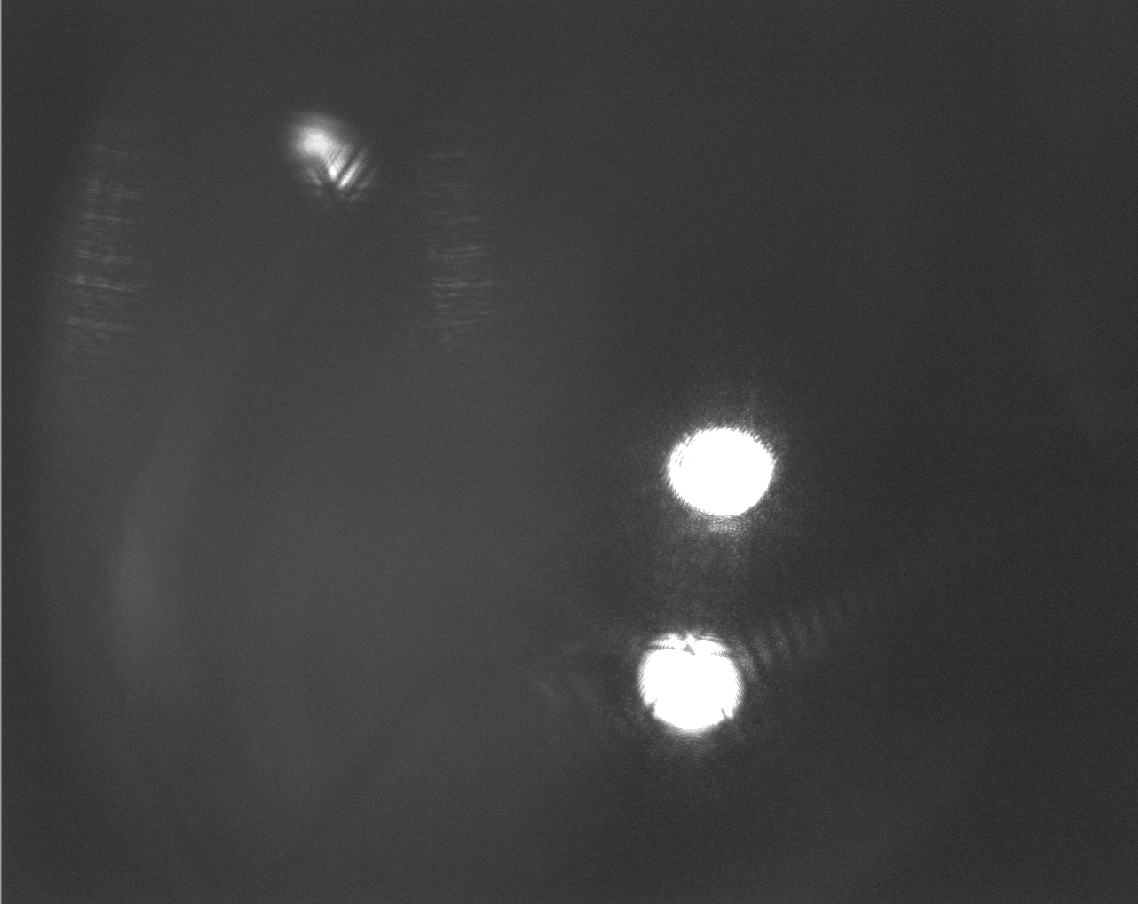}
	\caption{An Image Taken by the APATITE Camera Sensor Showing the Alignment Laser (Bottom) and Simulated Uplink Beacon (Top) Spots}
	\label{fig:APATITETestCamULAl}
\end{figure}

Figure~\ref{fig:APATITETestCamULAl} shows an image taken from the camera sensor of the alignment beam from JADE entering APATITE from the back of the module, and a simulated uplink beacon entering APATITE from the front. The beamsteering software will attempt to overlap the uplink beacon with the alignment beam, or align it to have a user-defined offset. A back-reflection of the alignment beam appears in the top-left of the image, for which there are fixes that can be applied to prevent errors from occurring in the beamsteering software: since the alignment beam is static, the back-reflection is also static, so an `ignore region' could be set. Otherwise, because the back-reflection is much dimmer than the two main beams, the camera sensor's exposure duration could be reduced so both of the bright beams can be detected but the back-reflection cannot. 

\begin{figure}
	\centering
	\includegraphics[width = \linewidth]{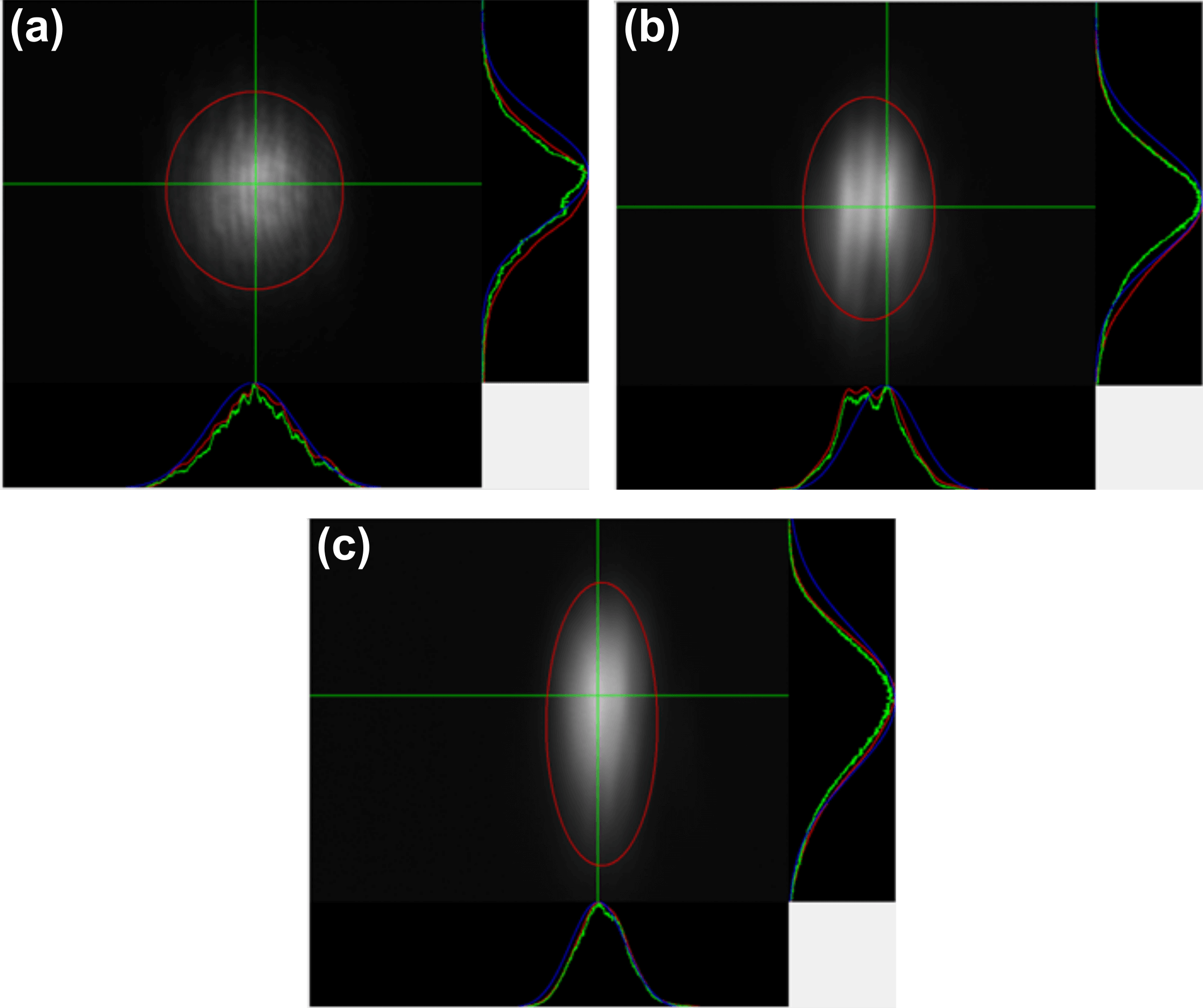}
	\caption{Quantum Source Beam Profile at (a) 20\,cm, (b) 120\,cm, and (c) 220\,cm from the Front of APATITE}
	\label{fig:APATITEBeamProfiles}
\end{figure}

\noindent \underline{Beam divergence measurements}: Requirements for APATITE state that the quantum source and downlink beacon laser beams should have full angle FWHM divergences before the telescope of 12\,$\upmu$rad and 3\,mrad, respectively. Divergence is measured by capturing images of a laser beam at several points along the beam propagation axis, using the distance along the beam and the beam width at various points to calculate the divergence. This technique also enables identification of beam astigmatism, and whether the beam has been truncated in the APATITE box due to misalignment. An example divergence measurement is shown in Fig.~\ref{fig:APATITEBeamProfiles} - this particular laser beam clipped the edge of an optic, causing an interference pattern most prominent 120\,cm away from APATITE (Fig.~\ref{fig:APATITEBeamProfiles}(b)), and causing the beam to converge on the x axis when it should diverge, most prominent 220\,cm away from APATITE (Fig.~\ref{fig:APATITEBeamProfiles}(c)). The most recently manufactured optomechanical mounts in APATITE provide correct alignment of all laser beams, eliminating this issue.

\begin{figure}
	\centering
	\includegraphics[width = \linewidth]{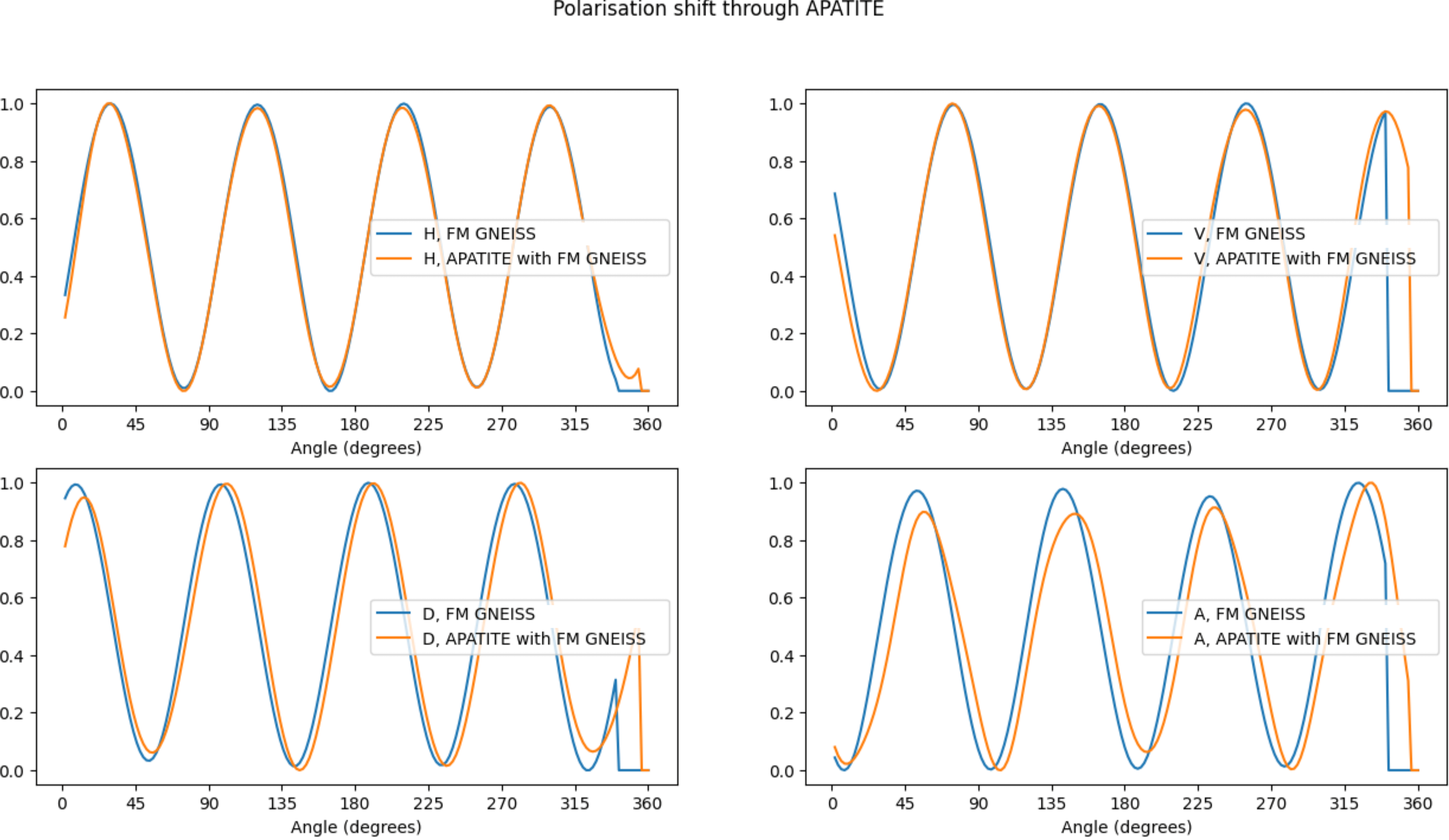}
	\caption{Relative Polarisation Shifts for Each of the Four Polarisation States After Passing Through APATITE}
	\label{fig:APATITEPolShift}
\end{figure}

\noindent \underline{Polarisation testing}: A relative polarisation test is performed with quantum source light generated by JADE and aligned through APATITE. Figure~\ref{fig:APATITEPolShift} shows the results of this test, overlapped with the results shown in Fig.~\ref{fig:JADERelativePolarisationMeasurement} (data smoothing has been applied for this figure to improve visibility of results). It was discovered that the earlier model of APATITE causes a rotation of $\sim8^\circ$ only in the D and A polarisations. From measurements performed on each of the optics individually and in combination, the root cause was identified to be the same optical clipping that induced interference in the beam profile seen in Fig.~\ref{fig:APATITEBeamProfiles}(b). 

\begin{figure}
	\centering
	\includegraphics[width = 5cm]{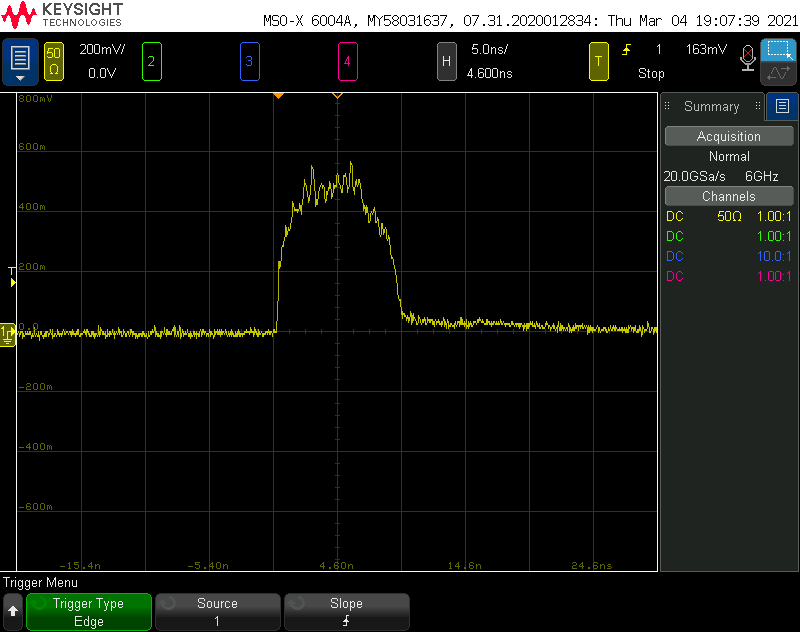}
	\caption{An Oscilloscope Measurement Showing the Optical Pulse Generated by the Downlink Beacon Laser in APATITE}
	\label{fig:DLBpulse}
\end{figure}

\noindent \underline{Downlink beacon testing}: The downlink beacon laser is pulsed for 10\,ns at a rate of 100\,kHz, so it operated on a 0.1\% duty cycle. An image of the optical pulse profile detected using a fast photodiode connected to an oscilloscope is shown in Fig.~\ref{fig:DLBpulse}. Using an optical power meter that averages the power readings, the measured power was 45\,mW - this corresponds to a peak power in the pulses of approximately 55\,W. 

\begin{figure}
	\centering
	\includegraphics[width = 5cm]{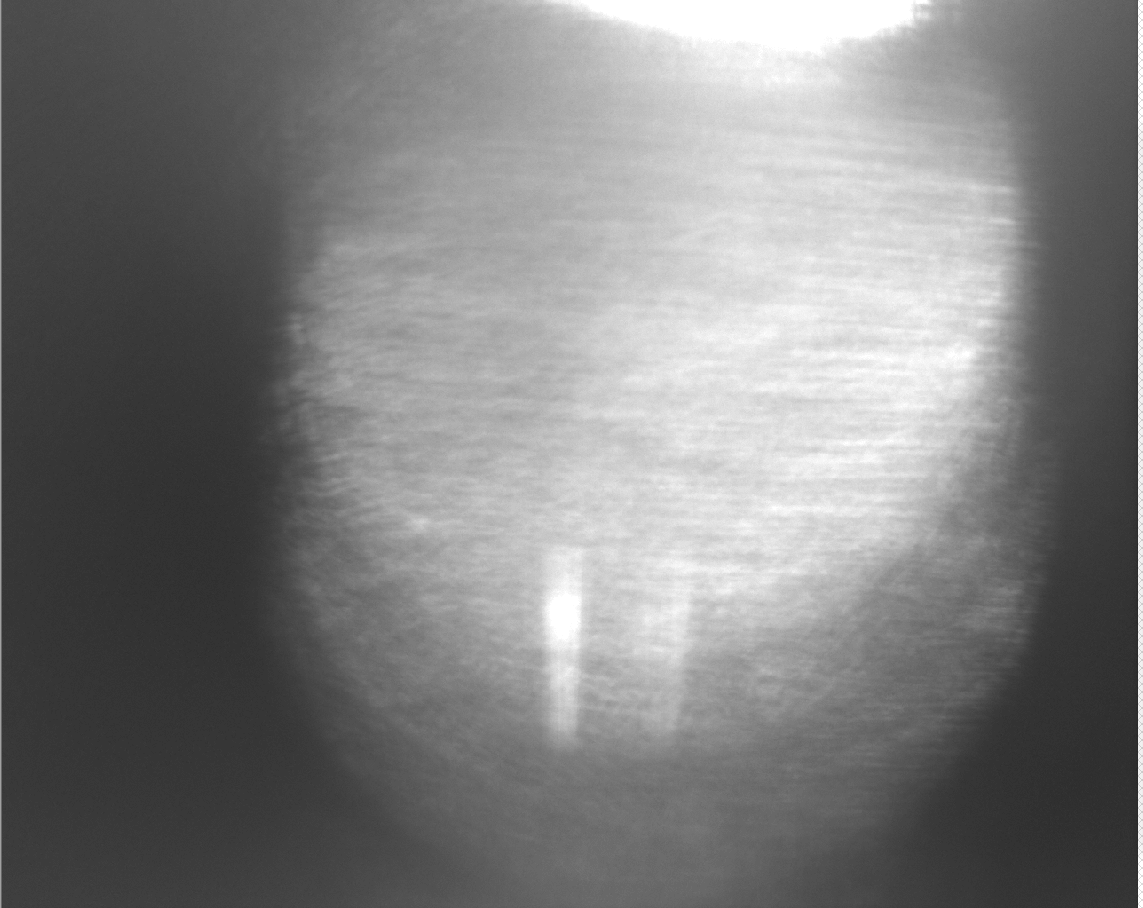}
	\caption{An Image Taken from the APATITE Camera Sensor While the Downlink Beacon was in Operation}
	\label{fig:DLBCameraBlinded}
\end{figure}

The APATITE camera sensor was operated during use of the downlink beacon, producing the image in Fig.~\ref{fig:DLBCameraBlinded}. Unfortunately the back-reflections caused by the downlink beacon beam would blind the camera sensor to the uplink beacon from the ground station. The most recent iteration of the mechanical part for which these back-reflections occur will be coated in Vantablack with the expectation to significantly reduce this effect.

\begin{figure}
	\centering
	\includegraphics[width = 5cm]{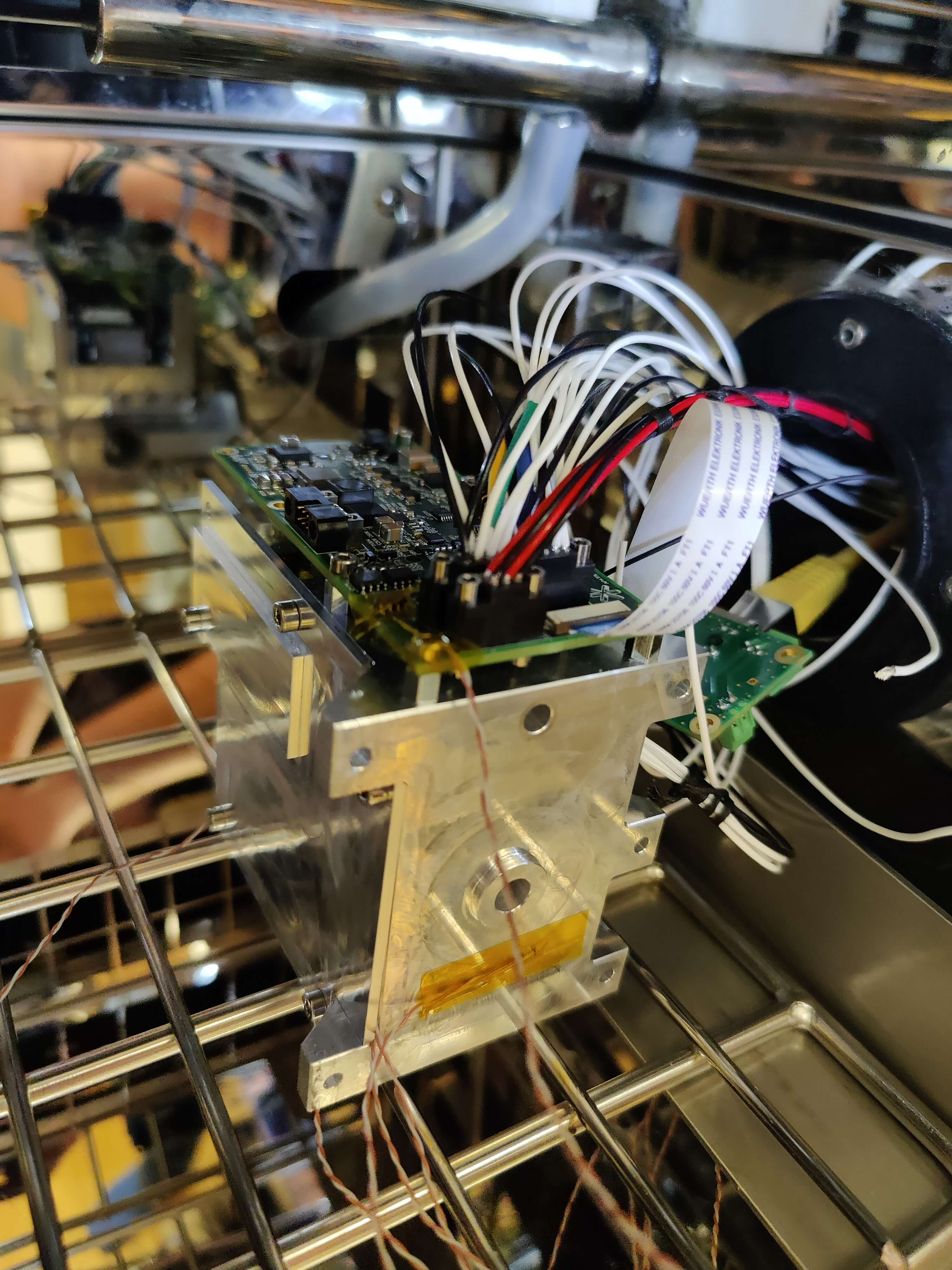}
	\caption{APATITE During Thermal Testing, All its Harnesses are Passed into a Feed-through Port in the Side of the Thermal Chamber}
	\label{fig:APAThermal}
\end{figure}

\noindent \underline{Thermal testing}: The APATITE camera sensor and MEMS mirror were tested at several points throughout the operating temperature range and after hot and cold starts at the extremes of this range. Both were found to operate normally without any issues under all the conditions experienced during the tests. An image of APATITE placed in the thermal chamber is shown in Fig.~\ref{fig:APAThermal}.

A visible test beam was attached to the fibre coupler to which JADE is usually connected, and aligned through the front window of the thermal chamber. The visible test beam was operated at room temperature, then at both of the extremes of the operating temperature range. Markers were placed to indicate where the beam was incident at each temperature allowing for the deflection angle to be measured. From 22$^\circ$C to 50$^\circ$C the beam was deflected by approximately 2.13\,mrad, and from 22$^\circ$C to -20$^\circ$C the beam was deflected by 1.81\,mrad in the opposite direction. It is presumed that these deflections were predominantly caused by thermal expansion and contraction of the aluminium parts as the temperature changed. In GARNET, the beams leaving APATITE will be magnified by a factor of 30 leading to a factor of 30 reduction in pointing errors, so over the full operating temperature range the resulting beam is expected to deflect by $\sim131\,\upmu$rad over the full operational temperature range. 

\subsection{GARNET Optical Telescope}
\begin{figure}
	\centering
	\includegraphics[width = \linewidth]{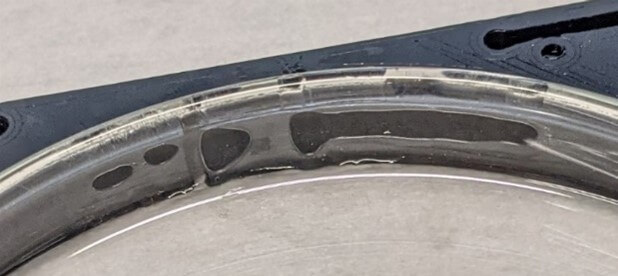}
	\caption{A Closeup Image of an Adhesive Test Performed at FHCAP}
	\label{fig:GARGlue}
\end{figure}

\noindent
Before the construction of GARNET, mechanical assemblies and bonding materials were characterised across the specified thermal range of ROKS, to confirm that the process of gluing components into their mounts would not lead to adverse effects on the optical performance of the telescope. A closeup of one of the test samples is shown in Fig.~\ref{fig:GARGlue}.

\begin{figure}
	\centering
	\includegraphics[width = \linewidth]{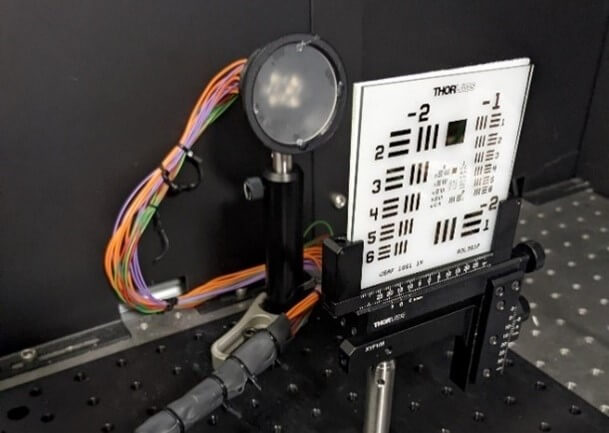}
	\caption{A Test Bench for GARNET Distortion Testing at FHCAP}
	\label{fig:GARTest}
\end{figure}

Once GARNET was assembled a test setup was used to ensure that the beam expansion by the telescope is uniform and that alignment of the telescope mirrors does not modify the mode. The test setup is shown in Fig.~\ref{fig:GARTest}, where some light sources are obscured by a frosted glass sheet. This assembly should provide uniform backlit illumination of the resolution test target, which can then be observed through the telescope. 

\begin{figure}
	\centering
	\includegraphics[width = \linewidth]{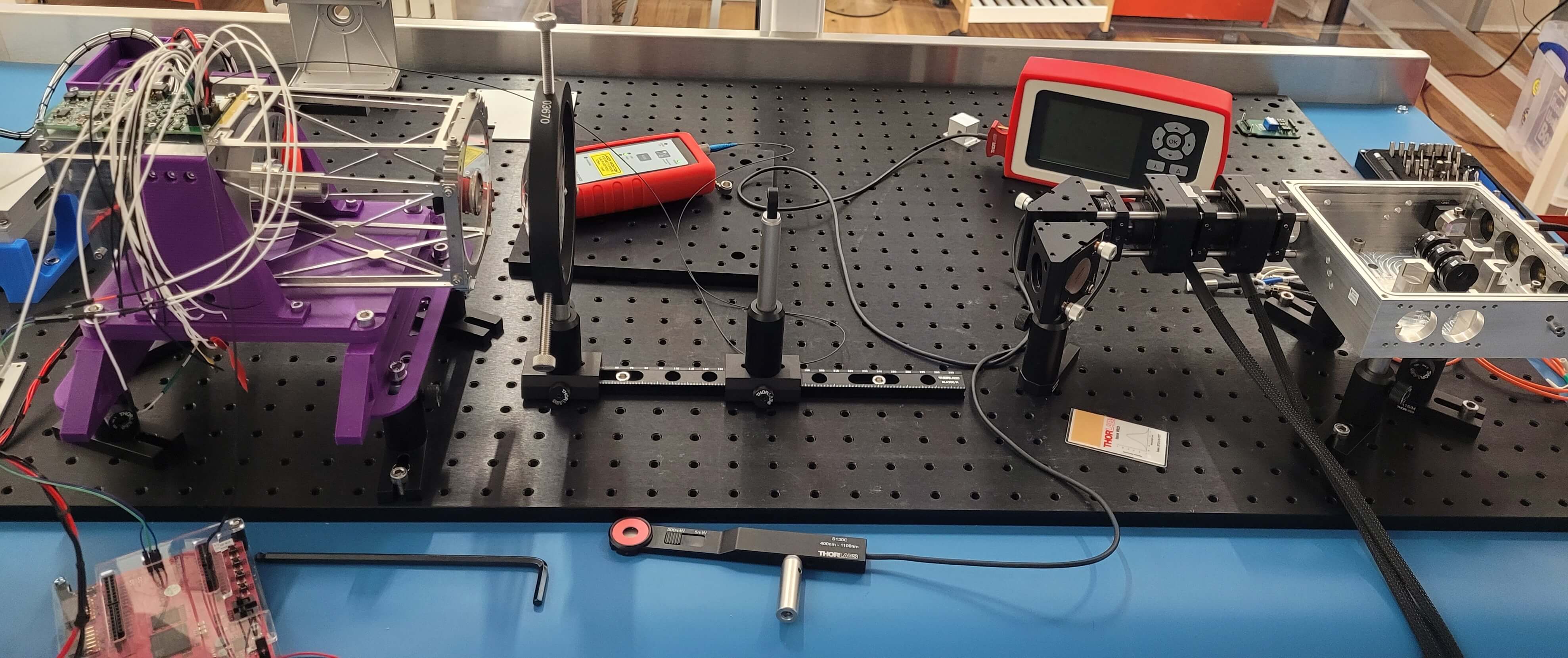}
	\caption{OPAL Polarisation Test Bench Early in the Build Process}
	\label{fig:OPALTestbench}
\end{figure}

Finally, once GARNET was delivered to CPL, APATITE was secured to the back of it and an OPAL test bench was constructed, as shown in Fig.~\ref{fig:OPALTestbench}. On the left is the OPAL assembly, on the right is the polarisation readout apparatus. A pair of lenses to shrink laser beam after GARNET has magnified them, can be seen in the centre. This test bench is being used to characterise the loss of optical power through the telescope due to a central obstruction required for the design, and will later be used to analyse the polarisations produced by the OPAL system from end-to-end.

\subsection{FLI-NT Intelligent Imaging Subsystem}
\begin{figure}
	\centering
	\includegraphics[width = \linewidth]{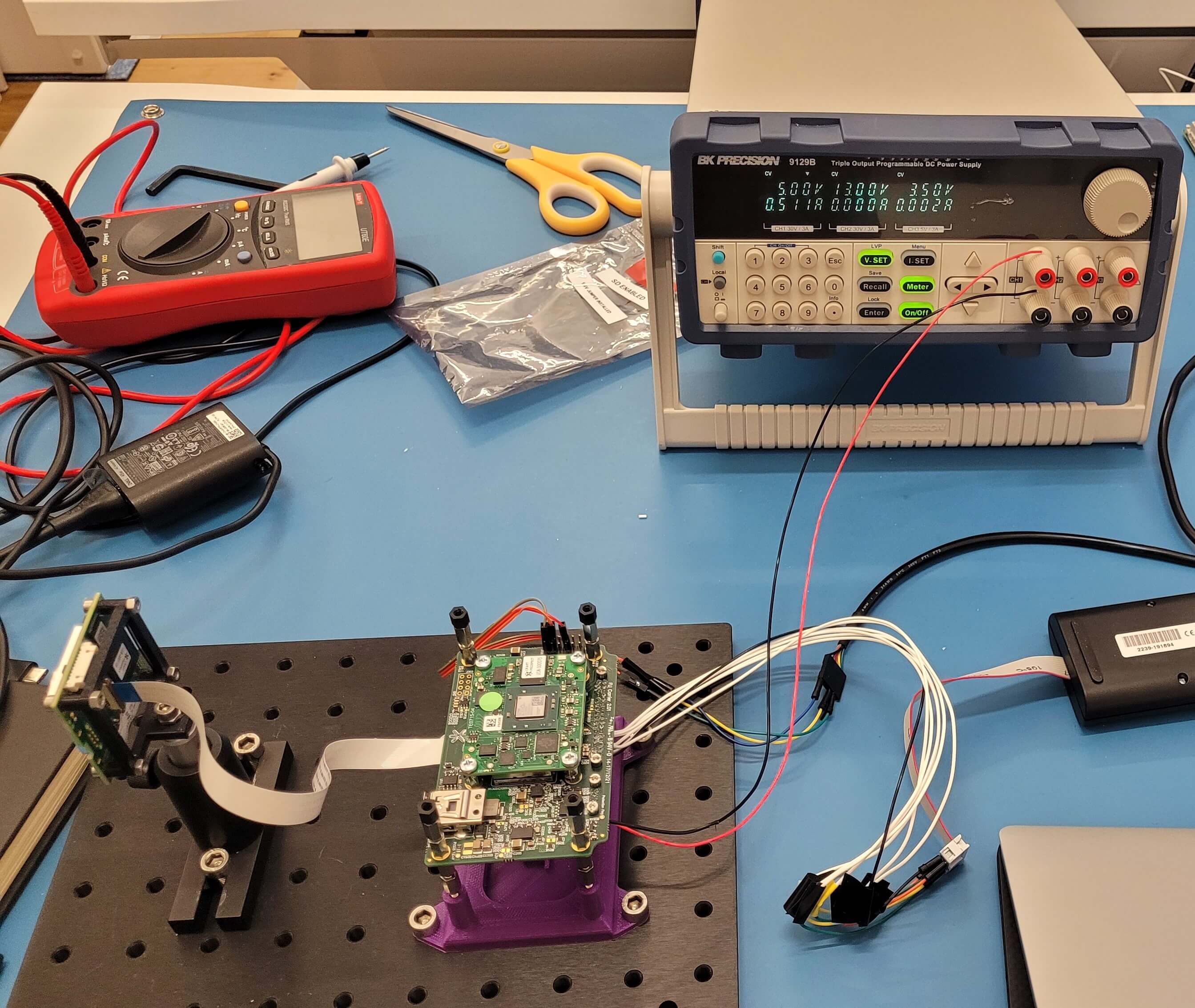}
	\caption{The Electronics for the FLINT Module Set Up on a Test Bench}
	\label{fig:FLINTTestbench}
\end{figure}

\noindent
Since the sole purpose of the FLI-NT module is to classify features in images, the baseline functional test after assembly is simply to record images. An example FLI-NT test bench that was used for this purpose can be seen in Fig.~\ref{fig:FLINTTestbench}. 

\begin{figure}
	\centering
	\includegraphics[width = \linewidth]{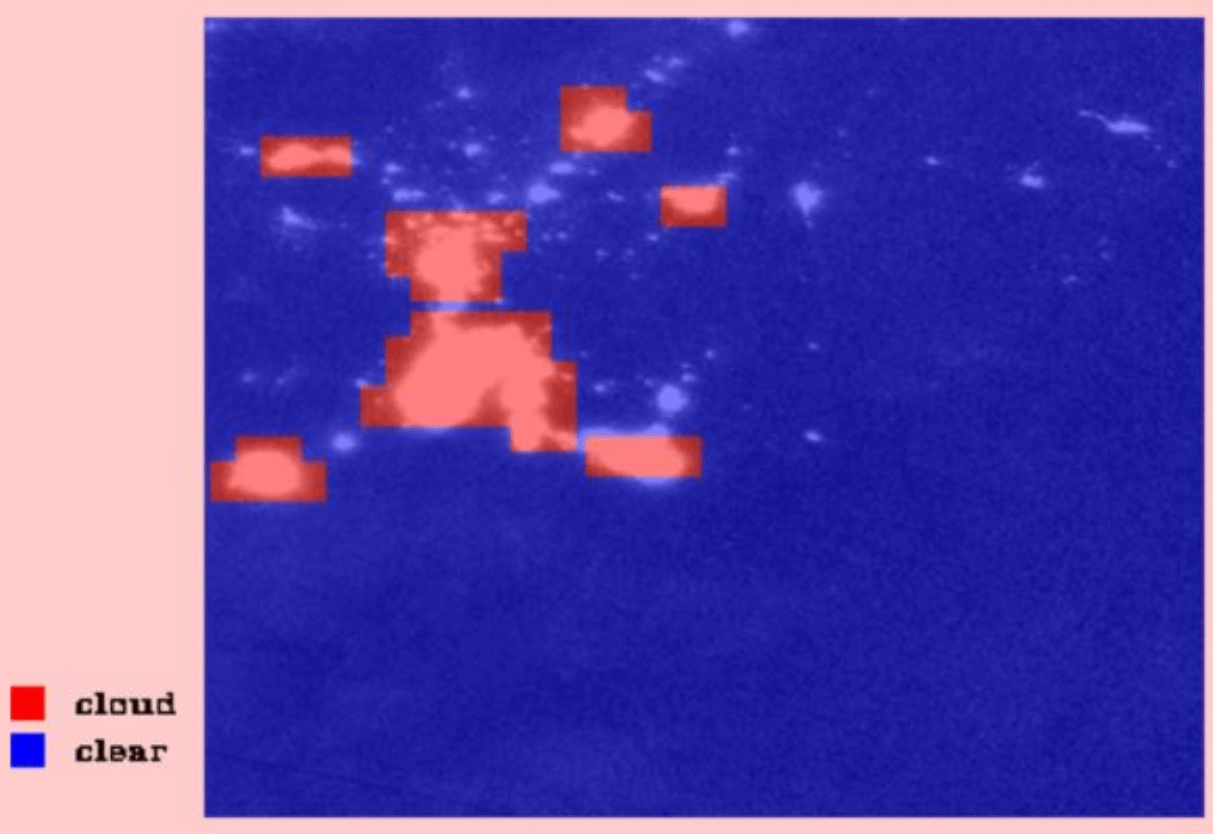}
	\caption{A Sample Cloud Image After Being Classified by FLI-NT During Functional Testing}
	\label{fig:FLINTTestImage}
\end{figure}

Images were successfully recorded during initial functional testing, at the extremes of the operational temperature range, and after hot and cold starts at these extremes. A sample cloud image that has been classified by a newly assembled FLI-NT module is shown in Fig.~\ref{fig:FLINTTestImage} - this variant of FLI uses a monochrome sensor optimised for the detection of infrared light, since it will be used to image clouds at night time, but false colour is added during the feature classification stage.

\subsection{OBSIDIAN Onboard Computer}
\begin{figure}
	\centering
	\includegraphics[width = \linewidth]{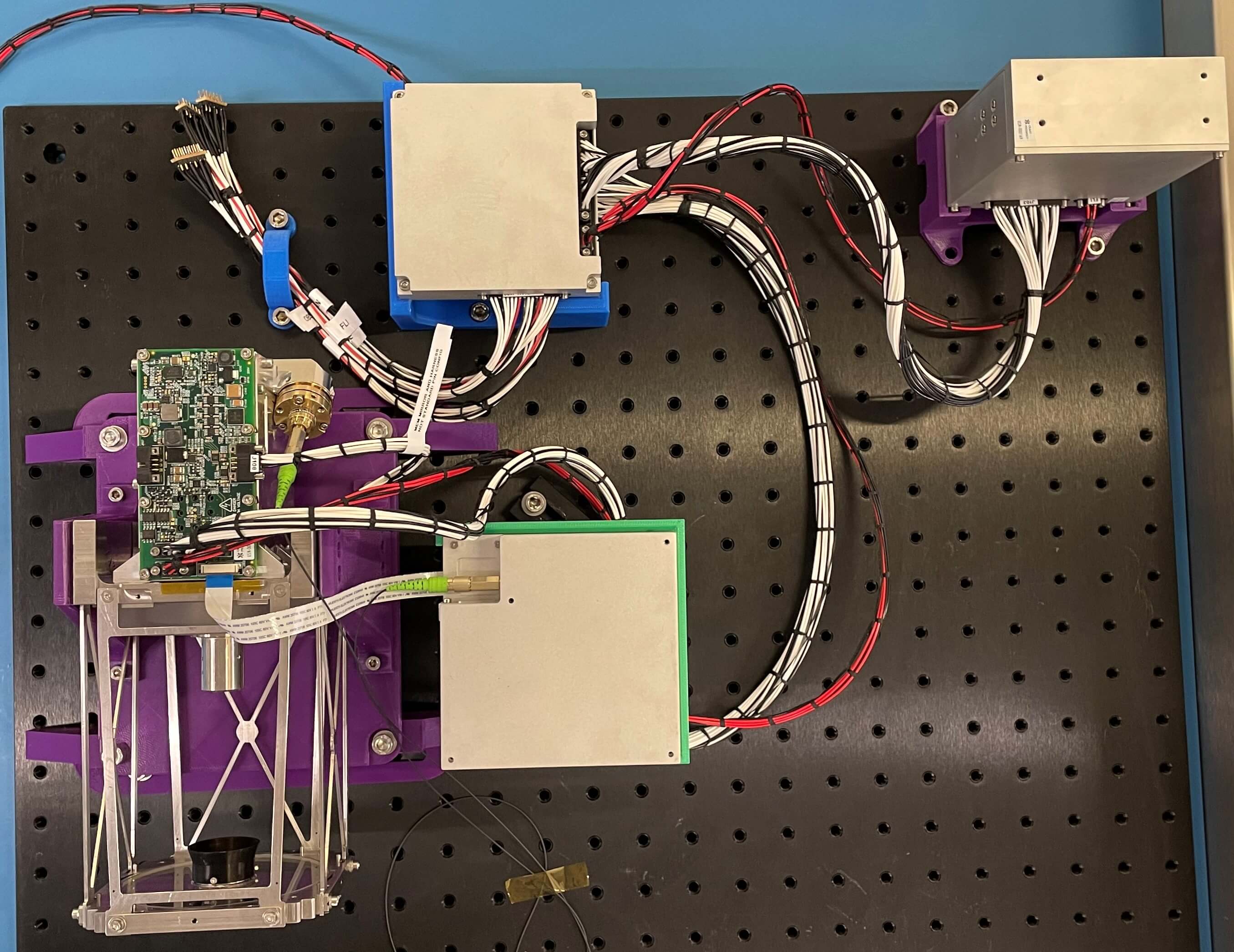}
	\caption{The Test Bench with All CPL Subsystems Connected for Integration Testing}
	\label{fig:OBSIDIANTestbench}
\end{figure}

\noindent
OBSIDIAN interfaces with all the other subsystems in ROKS, so initial tests involved integration with other modules on an individual basis to ensure OBSIDIAN could command them and capture telemetry data. After these tests were successfully executed, multiple subsystems were connected to OBSIDIAN and tested simultaneously. An image of one such test is shown in Fig.~\ref{fig:OBSIDIANTestbench}, where FLI-NT, JADE, and APATITE are electrically connected to OBSIDIAN. These tests also proved successful. 

Finally a thermal test was performed with JADE and FLI-NT connected to OBSIDIAN. It was demonstrated that all the modules worked together throughout the operational temperature range, even after hot and cold starts at the extremes of this range. 

\subsection{Platform Fit Checking}
\begin{figure}
	\centering
	\includegraphics[width = \linewidth]{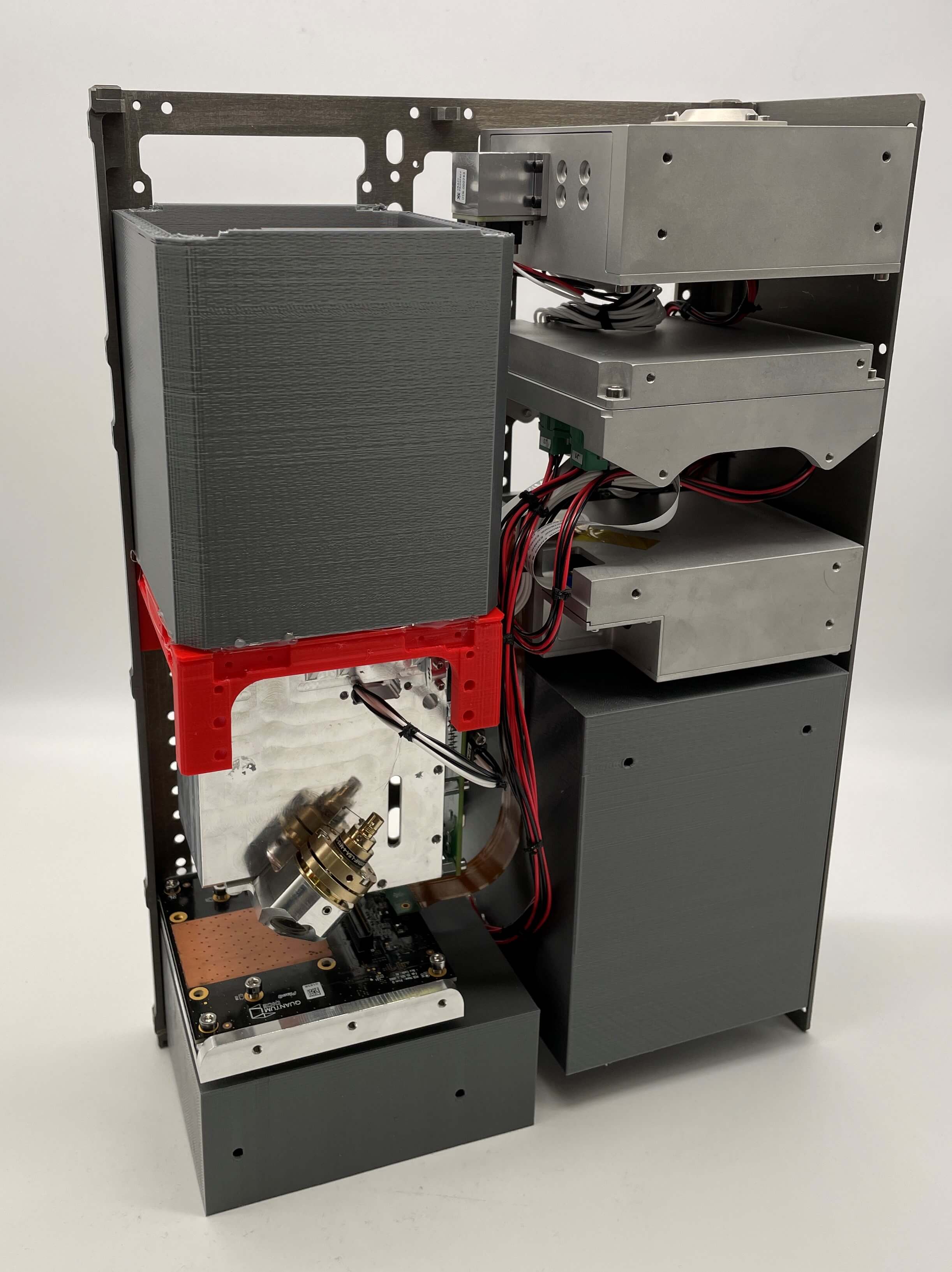}
	\caption{All the Prepared Modules Inside the 6U CubeSat Structure}
	\label{fig:FitCheck}
\end{figure}

\noindent
Once all the subsystems were confirmed working both individually and together, they were secured to a flight-representative 6U platform, shown in Fig.~\ref{fig:FitCheck}, to ensure that the modules and their harnesses would fit inside the CubeSat. Keep out zones and modules that await delivery are blocked out using 3D printed boxes of the same dimensions. Testing on the system as a whole in this form factor is imminent at the time of publication. 

\section{Optical Ground Station}
\label{sec:OGS}
\begin{figure*}
	\centering
	\includegraphics[width = \linewidth]{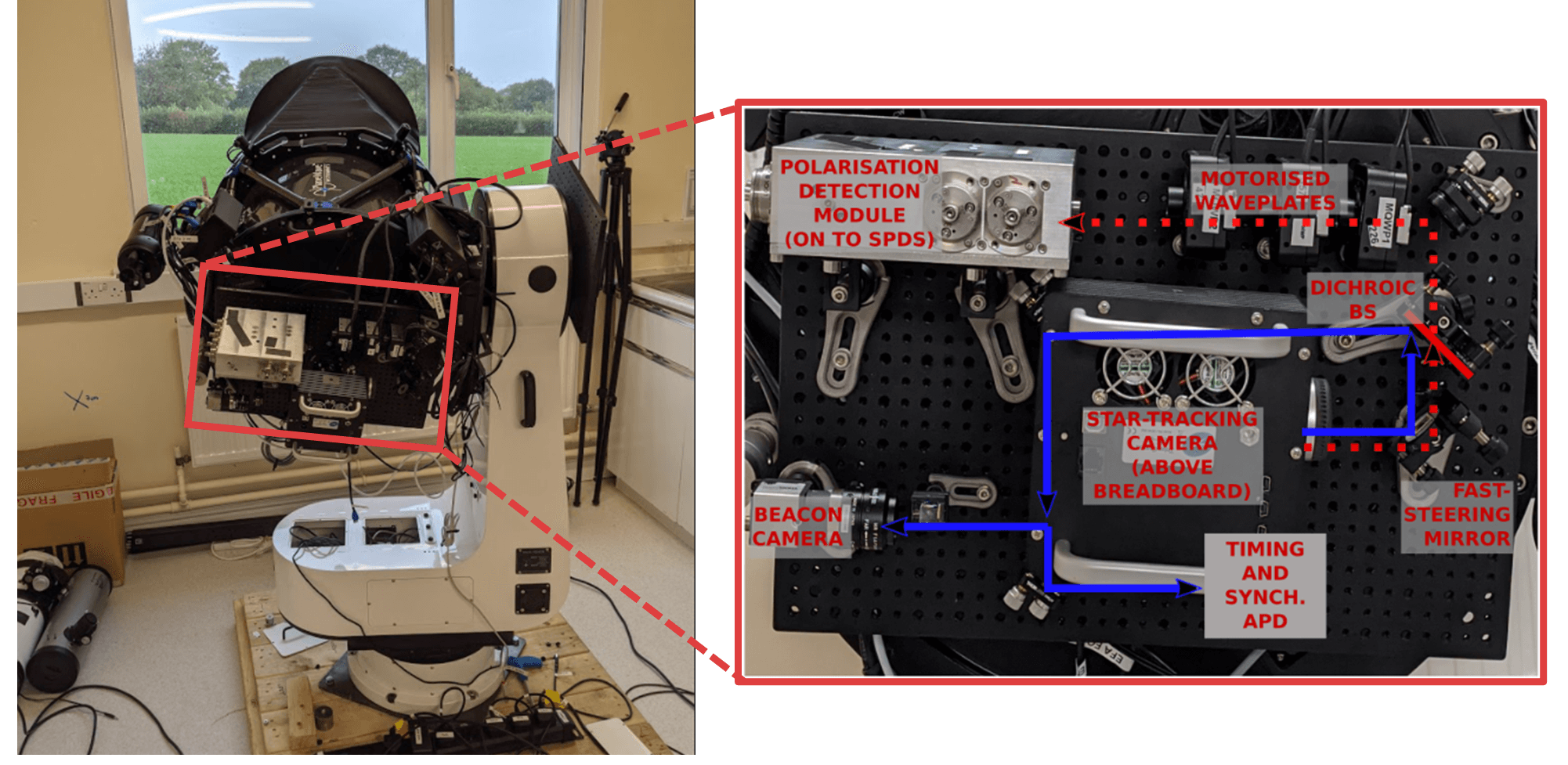}
	\caption{The Optical Ground Station Apparatus used at University of Bristol}
	\label{fig:Telescope}
\end{figure*}

\noindent
Mission partners at UoB are constructing and testing a portable OGS apparatus for ROKS. The OGS encompasses a commercial off-the-shelf telescope with custom receiver optics mounted on the back, as shown in Fig.~\ref{fig:Telescope}, and an optical breadboard where transmission optics can be affixed. Collaborative end-to-end QKD tests were initially planned between CPL and UoB for this phase of ROKS, however COVID-19 posed significant challenges for integrated testing and in-person visits. The team worked around these issues by constructing apparatus at CPL for benchtop QKD demonstrations, as seen in Fig.~\ref{fig:OPALTestbench}, and by setting up experiments with the telescope at UoB using downlink beacon representative lasers rather than a quantum source. The experiments at both facilities were successful, and integrated end-to-end testing is still anticipated.

\section{LESSONS LEARNED}
\label{sec:Lessons}
\noindent
The work presented in the previous sections was the culmination of an intense eight month development period that saw engineering models of numerous products advance to flight representative models and integrated into a single payload. What would already be a significant undertaking for a relatively small team was made more challenging by partial lockdowns due to COVID-19, an ongoing chip shortage leading to part sourcing and stockpiling, and longer than usual lead times for components generally. 

Furthermore, it was found upon receipt that some optics did not meet the needs of the mission, and in some cases did not meet the supplier specifications. Optics are particularly challenging for this project given the unusual requirement that relationships between four different polarisations must be maintained across several optical components. Clarity of what is expected of suppliers and their products from the outset, an improved goods-in procedure, and enhanced testing capabilities became critical to avoiding similar issues with suppliers.

The ROKS team is multidisciplinary, and no one person has an intimate knowledge of all aspects of any system. Test documentation was crucial to reduce the people-hours spent on testing and troubleshooting, and to ensure there were no single points of failure. This prevented team members from being blocked from completing tests of troubleshooting procedures when they came across aspects of mechanical or optical hardware, electronics, or software with which they were unfamiliar, lessening the impact of team absences.

Many of the lessons learned during this phase of the project were due to an inexperienced team facing its first extensive integration period. Many of the issues encountered were normal integration challenges, and implementing or tweaking procedures on-the-fly led to a reduction of issues or improvements to the outcomes as time went on. A `lessons learned' session was organised with the team at CPL to discuss and identify areas for improvement, to ease MAIT stages in future projects. The key messages were as follows:

\begin{itemize}
	\item Have spares of every part or component delivered
	\item Photograph or video every process
	\item Serial number all parts and assemblies
	\item Keep record cards with modules and assemblies
	\item Record test configurations
	\item Invest in test setups
	\item Maintain test setups and keep them tidy
	\item Automate repeatable functional tests
	\item Start end-to-end testing early
	\item Prototype early and often	
	\item Check supplied parts against their specs
	\item Plan testing to an appropriate level before executing
	\item Design parts and assemblies for handling
	\item Communicate continuously throughout the team
	\item Plan for forgetfulness and thoughtlessness
	\item Verify and update budgets frequently
	\item Record steps taken in detail
	\item Plan for subcontractor production issues
	\item Plan for multiple models
	\item Speak up if work becomes overwhelming
\end{itemize}

\section{CONCLUSIONS}
\label{sec:Conclusions}
\noindent
The progress made on ROKS payload development represents a significant milestone for Craft Prospect Ltd and has resulted in the production and integration of several flight modules for a range of subsystems. These modules will be made available as configurable products across quantum technology and onboard intelligence. Lessons learned and shared throughout the team will be invaluable in extensive future missions that will span from feasibility studies through to payload and system delivery. 

The baseline results in Section~\ref{sec:ModuleTesting} show that the flight models behave as expected throughout the temperature range anticipated in LEO. These flight models are subject to continuous tweaks and improvements to ensure the success and longevity of the ROKS mission. The integration and testing work will continue, full payload thermal and vibe will ensue once the platform bus system has been delivered, and a flight readiness review is forecast to occur by the end of 2022. 

\section{ACKNOWLEDGEMENTS}
\noindent
The ROKS team would like to extend its appreciation to all partners and team members past and present. In particular CPL acknowledge the support and expertise of Fraunhofer Research UK, Bright Ascension Ltd, Orbital Astronautics Ltd, University of Strathclyde, and University of Bristol, with support from ESA, Innovate UK, UK Space Agency, Quantum Communications Hub, and all private contributors and industrial supporters of the mission.

The authors would like to thank the UK Space Agency for the continued funding (National Space Innovation Program NSIP-07) and support that facilitated this work.

\bibliographystyle{unsrt}

\end{document}